\documentclass[11pt]{article}

\usepackage[square,numbers,sort&compress]{natbib}
\usepackage{graphicx}
\usepackage{amsmath,amssymb}
\usepackage{authblk}
\usepackage[margin=1in]{geometry}
\usepackage{hyperref}

\begin{document}

\title{Deep learning approaches for neural decoding: from CNNs to LSTMs and spikes to fMRI}
\author[1,2,*]{Jesse A. Livezey}
\author[3,4,5,*]{Joshua I. Glaser}
\affil[ ]{\href{jlivezey@lbl.gov}{jlivezey@lbl.gov}, \href{j.glaser@columbia.edu}{j.glaser@columbia.edu}}
\affil[*]{equal contribution}
\affil[1]{Biological Systems and Engineering Division, Lawrence Berkeley National Laboratory, Berkeley, California, United States}
\affil[2]{Redwood Center for Theoretical Neuroscience, University of California, Berkeley, Berkeley, California, United States}
\affil[3]{Department of Statistics, Columbia University, New York, United States}
\affil[4]{Zuckerman Mind Brain Behavior Institute, Columbia University, New York, United States}
\affil[5]{Center for Theoretical Neuroscience, Columbia University, New York, United States}

\maketitle

\abstract{
Decoding behavior, perception, or cognitive state directly from neural signals has applications in brain-computer interface research as well as implications for systems neuroscience. In the last decade, deep learning has become the state-of-the-art method in many machine learning tasks ranging from speech recognition to image segmentation. The success of deep networks in other domains has led to a new wave of applications in neuroscience. In this article, we review deep learning approaches to neural decoding. We describe the architectures used for extracting useful features from neural recording modalities ranging from spikes to EEG. Furthermore, we explore how deep learning has been leveraged to predict common outputs including movement, speech, and vision, with a focus on how pretrained deep networks can be incorporated as priors for complex decoding targets like acoustic speech or images. Deep learning has been shown to be a useful tool for improving the accuracy and flexibility of neural decoding across a wide range of tasks, and we point out areas for future scientific development.}

\section{Introduction}

Using signals from the brain to make predictions about behavior, perception, or cognitive state, i.e., ``neural decoding", is becoming increasingly important within neuroscience and engineering. One common goal of neural decoding is to create brain computer interfaces, where neural signals are used to control an output in real time. This could allow patients with neurological or motor diseases or injuries to, for example, control a robotic arm or cursor on a screen, or produce speech through a synthesizer. Another common goal of neural decoding is to gain a better scientific understanding of the link between neural activity and the outside world. To provide insight, decoding accuracy can be compared across brain regions, cell types, different types of subjects (e.g., with different diseases or genetics), and different experimental conditions~\cite{quiroga2006movement, harrison2009decoding, acharya2010electrocorticographic, weygandt2012fmri, rich2016decoding, glaser2018population, hamilton2018spatial, brackbill2020reconstruction}. Plus, the representations learned by neural decoders can be probed to better understand the structure of neural computation~\cite{mcintosh2016deep, nagamine2017understanding, kell2018task, livezey2019deep}. These uses of neural decoding span many different neural recording modalities and span a wide range of behavioral outputs (Fig.~\ref{fig:1}A).

Within the last decade, many researchers have begun to successfully use deep learning approaches for neural decoding. A decoder can be thought of as a function approximator, doing either regression or classification depending on whether the output is a continuous or categorical variable. Given the great successes of deep learning at learning complex functions across many domains~\cite{alipanahi2015predicting, piech2015deep, paganini2018calogan, kurth2018exascale, schutt2018schnet,hochreiter1997long, krizhevsky2012imagenet, sutskever2014sequence, he2016deep, amodei2016deep}, it is unsurprising that deep learning has become a popular approach in neuroscience. Here, we will review the many uses of deep learning for neural decoding. We will emphasize how different deep learning architectures can induce  biases that can be beneficial when decoding from different neural recording modalities and when decoding different behavioral outputs. We hope this will prove useful to deep learning researchers aiming to understand current neural decoding problems and to neuroscience researchers aiming to understand the state-of-the-art in neural decoding.

\section{Deep learning architectures}\label{sec:architectures}

At their core, deep learning models share a common structure across architectures: 1) simple components formed from linear operations (typically matrix multiplication or convolution) plus a nonlinear operation (for example, rectification or a sigmoid nonlinearity); and 2) composition of these simple components to form complex, layered architectures. There are many formats of neural networks, each with their own set of assumptions. In addition to feedforward neural networks, which have the basic structure described above, common architectures  for neural decoding are convolutional neural networks (CNNs) and recurrent neural networks (RNNs). While more complex deep network layer types, e.g., graph neural networks~\cite{wu2020comprehensive} or networks that use attention mechanisms~\cite{vaswani2017attention}, have been developed, they have not seen as much use in neuroscience. Additionally, given that datasets in neuroscience typically have limited numbers of trials, simpler, more shallow deep networks (e.g., a standard convolutional network versus a residual convolutional network \cite{he2016deep}) are often used for neural decoding.

\begin{figure}[!tpb]%figure1
\centering
\includegraphics[width=\textwidth]{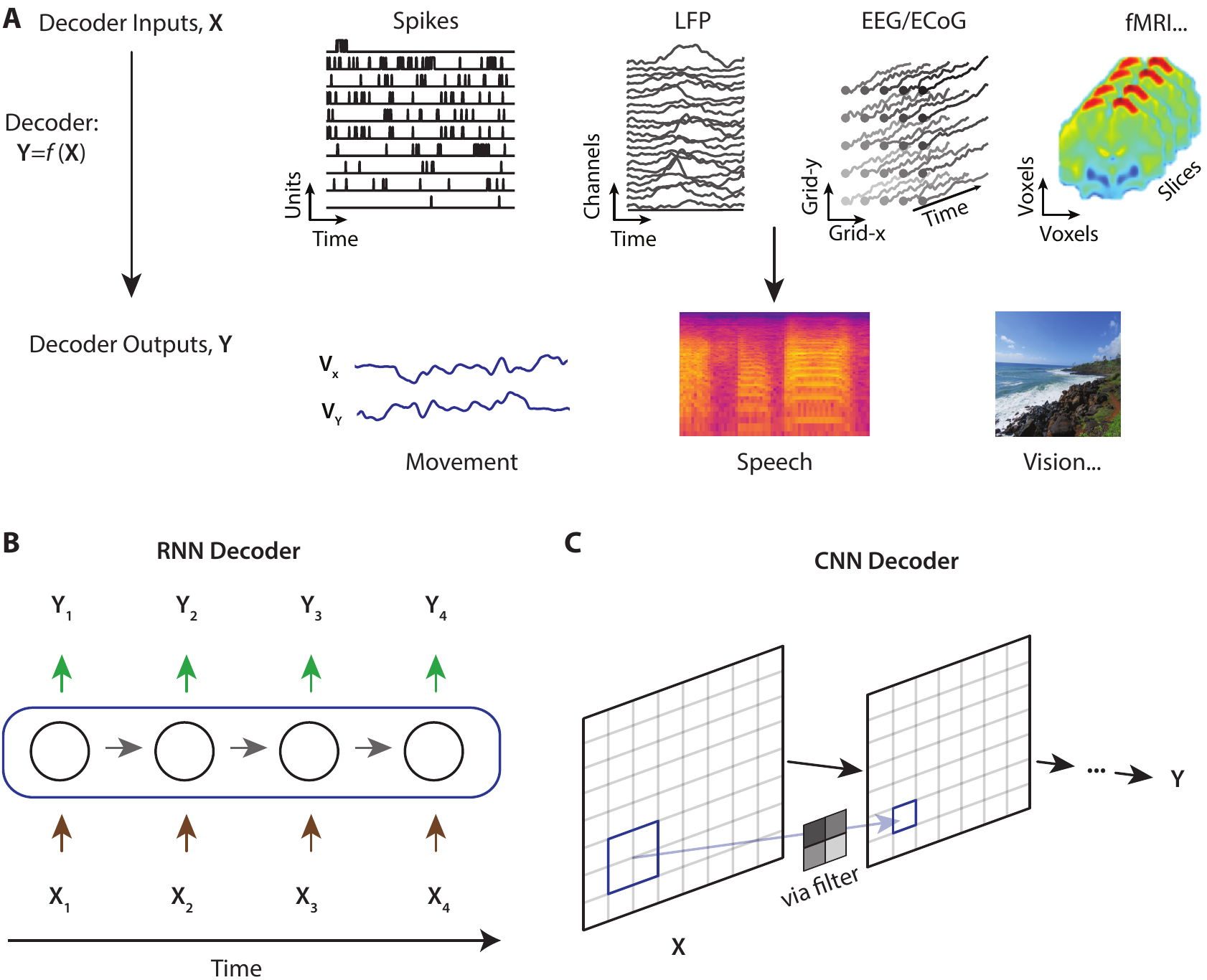}
\caption{Schematics. {\bf A:} Schematics of neural decoding, which can use many different neural modalities as input (top) and can predict many different outputs (bottom). Embedded figures are adapted from  ~\cite{glaser2017machine,bouchard_chang_2019,kazemifar2017spontaneous}. {\bf B:} A schematic of a standard recurrent neural network (RNN). Each arrow represents a linear transformation followed by a nonlinearity. Arrows of the same color represent the same transformations occurring. The circles representing the hidden layer typically contain many hidden units. More sophisticated versions of RNNs, which include gates that control information flow through various parts of the network, are commonly used. For example, see \cite{goodfellow2016deep} for a schematic of an LSTM. {\bf C:} A schematic of a convolutional neural network. A convolutional transformation takes a learned filter and convolves it with the input (here, a 2d input), and then passes this through a nonlinearity. This means that here, a 2$\times$2 filter will be multiplied pixel-wise with all 2$\times$2 blocks to get the values of the next layer in the network.
} \label{fig:1}
\end{figure}

RNNs typically use a sequence of inputs. RNNs are also capable of processing inputs that are sequences of varying lengths, which occurs in neuroscience data (e.g., trials of differing duration). This is unlike a fully-connected network, which requires a fixed dimensionality input. In an RNN, the inputs are then projected into a hidden layer, which connects to itself across time (Fig.~\ref{fig:1}B). Thus, recurrent networks are commonly used for decoding since they can flexibly incorporate information across time. Finally, the hidden layer projects to an output, which can itself be a sequence (Fig.~\ref{fig:1}B), or just a single data point. 

CNNs can be adapted to input and output data in many different formats. For example, convolutional architectures can take in structured data (1d timeseries, 2d images, 3d volumes) of arbitrary size. The convolutional layers will then learn filters of the corresponding dimensions, in order to extract meaningful local structure (Fig.~\ref{fig:1}C). The convolutional layers will be particularly useful if there are important features that are translation invariant, as in images. This is done hierarchically, in order to learn filters of varying scales (i.e., varying temporal or spatial frequency content). Next, depending on the output that is being predicted, the convolutional layers are fed into other types of layers to produce the final output (e.g., into fully connected layers to classify an image). In general, hierarchically combining local features is a useful prior for image-like datasets.

Weight-sharing, where the weights of some parameters are constrained to be the same, is often used for neural decoding. For instance, the parameters of a convolutional (in time) layer can be made the same for differing input channels or neurons, so that these inputs are filtered in the same way. This is analogous to CNN parameters being shared across space or time in 2d or 1d convolutions. For neural decoding, this can be beneficial for learning a shared set of data-driven features for different recording channels as an alternative to human-engineered features.

Training a neural decoder uses supervised learning, where the network's parameters are learned to predict target outputs based on the inputs. Recent work has combined supervised deep networks with unsupervised learning techniques. These unsupervised methods learn (typically) lower dimensional representations that reproduce one data source (either the input or output), and are especially prevalent when decoding images. One common method, generative adversarial networks (GANs)  ~\cite{goodfellow2014generative, radford2015unsupervised}, generate an output, e.g. an image, given a vector of noise as input. GANs are trained to produce images that fool a classifier deep network about whether they are real versus generated images. Another method is convolutional autoencoders, which are trained to encode an image into a latent state, and then reconstruct a high fidelity version~\cite{parthasarathy2017neural}. These unsupervised methods can produce representations of the decoding input or output that are sometimes more conducive for decoding.

\section{The inputs of decoding: neural recording modalities and feature engineering}

\subsection{Neural recording modalities}

To understand how varying neural network architectures can be preferable for processing different neural signals, it is important to understand the basics of neural recording modalities. These modalities differ in their invasiveness, and their spatial and temporal precision.

The most invasive recordings involve inserting electrodes into the brain to record voltages. This allows experimentalists to record \textit{spikes} or \textit{action potentials}, the fast electrical transients that individual neurons use to signal, and the basic unit of neural signaling. To get binary spiking events, the recorded signals are high-pass filtered and thresholded. Datasets with spikes are thus binary time courses from all of the recording channels (Fig.~\ref{fig:1}A). These invasive measurements also allow recording \textit{local field potentials} (\textit{LFPs}), which are the low-pass filtered version (typically below $\sim$200Hz) of the same recorded voltage. LFPs are thought to be the sum of input activity of local neurons~\cite{buzsaki2012origin}. When all voltage is included across frequency bands, the voltage is generally referred to as \textit{wide-band activity}. Datasets with LFP and wide-band are continuous time courses of voltages from all the recording channels (Fig.~\ref{fig:1}A). Note that traditionally, due to the distance between recording electrodes being greater than the spatial precision of recording, spatial relationships between electrodes are not utilized for decoding. Spikes, LFP, and wide-band are more commonly recorded from animal models than humans because of their invasive nature.

Another invasive technique for recording individual neurons' activities is \textit{calcium imaging}, which uses microscopy to capture images of fluorescent calcium indicators that are sensitive to neurons' spiking activity \cite{chen2013ultrasensitive}. The raw outputs of calcium imaging are videos: pixels measure fluorescence at the times when, and locations where, neurons are active. Calcium imaging is only used with animal models.

Electrical potentials measured from outside of the brain, that is \textit{electrocorticography} (\textit{ECoG}) and \textit{electroencephalography} (\textit{EEG}), are common neural recording modalities used in humans. ECoG recordings are from grids that record electrical potentials from the surface of the cortex, require surgical implantation, and often cover large function areas of cortex. EEG is a noninvasive method that records from the surface of the scalp from up to hundreds of spatially distributed channels. Like LFPs, datasets from ECoG and EEG recordings are continuous time courses of electrical potentials across recording channels (Fig.~\ref{fig:1}A), but here the spatial layout of the channels is also sometimes used in decoding. Note that as these electrical recording methods get less invasive, spatial precision decreases (from spikes to LFP to ECoG to EEG), which can lead to inferior decoding performance \cite{flint2012local,frey2019deepinsight}. Still, all these electrical signals can be recorded at high temporal resolution (100s-1000s of Hz) which make them good candidates for fast time-scale decoding.

\textit{Magnetoencephalography} (\textit{MEG}),  \textit{functional near infrared spectroscopy} (\textit{fNIRS}), and \textit{functional magnetic resonance imaging} (\textit{fMRI}) are also noninvasive recording modalities which are most often used in human decoding experiments. MEG measures the weak magnetic fields that are induced by electrical currents in the brain. Like EEG and ECoG, MEG can be recorded with high temporal precision. fNIRS and fMRI measure blood oxygenation (a proxy for neural activity), through its absorption of light and with resonance imaging respectively, and their temporal resolution are temporally limited by its dynamics. fNIRS and fMRI datasets contain activity signals in different ``voxels” (locations) of the brain over time. Due to the limited temporal resolution, sometimes the temporal continuity of this data is not used for decoding purposes (Fig.~\ref{fig:1}A).

\subsection{Feature engineering}

\begin{figure}[!tpb]%figure2
\centering
\includegraphics[width=\textwidth]{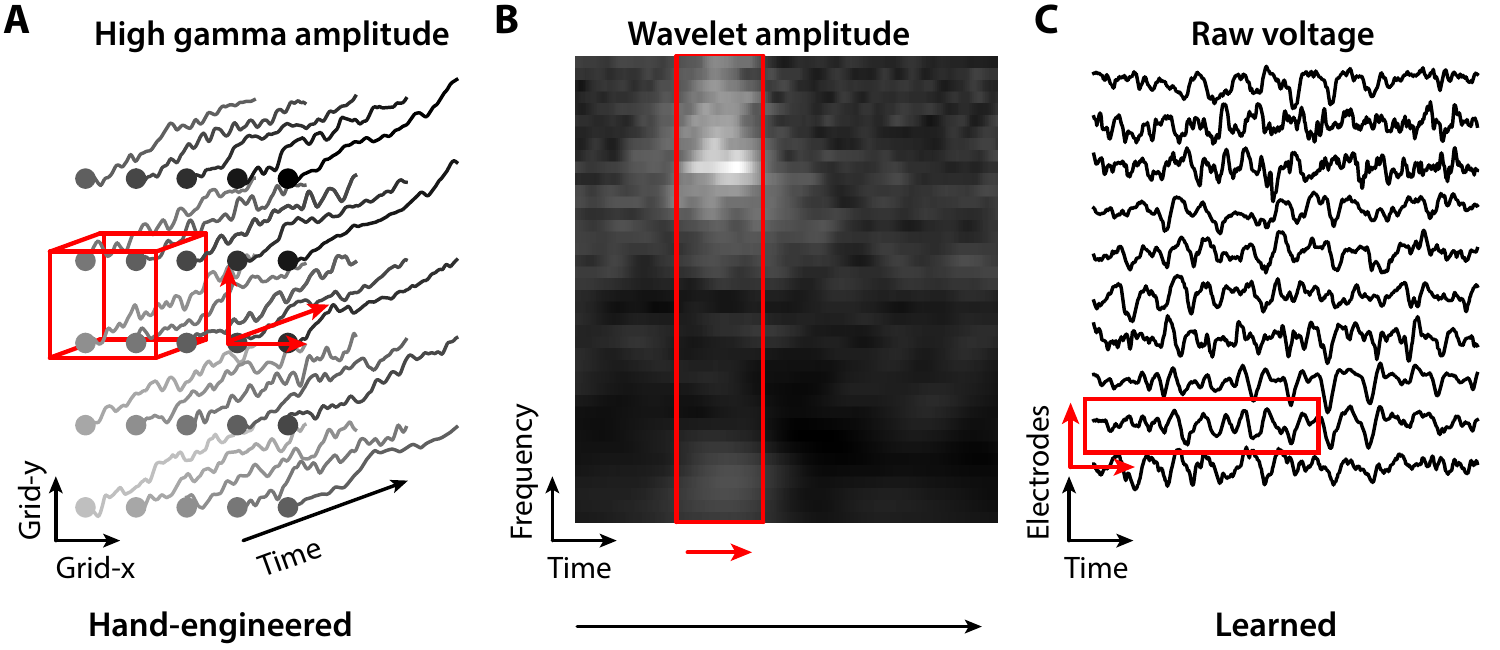}
\caption{Feature engineering for neural decoding. For all plots, the red box indicates a set of features across time, space, or frequency which will be filtered together by the first layer's convolutional or recurrent window. The red arrows indicate axes along which convolution or recurrence may be performed. Sample data from~\cite{bouchard_chang_2019}. {\bf A:} High gamma amplitude, which is selected from a large filterbank of features from {\bf B}, is shown spatially laid out in the ECoG grid locations. Deep network filters combine hand-engineered high gamma features across space and time. {\bf B:} Spectrotemporal wavelet decomposition of the raw data, from {\bf C}, may be used as the input to a deep network. The deep network filter shown combines features across frequency and time and can be shared across channels. {\bf C:} Raw electrical potential recorded using ECoG across channels. The deep network filter shown combines features across time and can be shared across channels.}\label{fig:2}
\end{figure}

For each of these recording modalities, the raw data are processed to create features that are beneficial for decoding. Sometimes, these features are hand-engineered based on previous knowledge, traditionally with the goal of creating features that are most compatible with linear decoders. Other times, this feature engineering is part of the deep learning architecture. That is, a more raw form of the input is provided into the decoder, and a first stage of the deep network decoder will automatically learn to extract relevant features. Specific neural network architectures can be beneficial for this automatic feature engineering (Fig.~\ref{fig:2}).

For use in decoding, spikes are typically first converted into firing rates by determining the number of spikes in time bins. Then, these firing rates are fed into the decoder. This general approach of decoding based on firing rates (an assumption of ``rate coding") is standard. While using precise temporal timing of spikes (``temporal coding") for decoding has been done \cite{maia2013functional}, we are not aware of examples using deep learning. Given that firing rates are used as inputs, additional neural network architectures are not used to extract unknown features from the input. However, in future research, it might be advantageous to provide a more raw form of spiking as input, and use deep learning architectures to do feature engineering. For rate coding, the best size and temporal placement of time bins could be automatically determined, and for temporal coding, features related to the precise timing of spikes could be learned.

When analyzing calcium imaging data, the videos are typically preprocessed to extract time traces of fluorescences over time for each neuron \cite{giovannucci2019caiman}. Sometimes, additional processing will be done to estimate spiking events from the calcium traces \cite{vogelstein2010fast}. Deep learning tools exist for both of these processing steps \cite{soltanian2019fast,speiser2017fast}. For decoding, either the fluorescences, or the estimated firing rates (via the estimated spike trains), are then used as input. While it could be possible to develop an end-to-end decoder that works with the videos as input, this may prove challenging given the potential for overfitting with high-dimensional input.

When decoding from wide-band, LFP, EEG, and ECoG data, it is common to first extract spectrotemporal features from the data, for example the signals in specific frequency bands. Sometimes, only ``task-relevant" frequencies will be used for decoding - for instance, using high gamma frequencies in ECoG to decode speech~\cite{bouchard2014neural, yang2015speech} (Fig.~\ref{fig:2}A). More frequently, many frequencies will be included, to better understand which are contributing to decoding \cite{mugler2014direct, livezey2019deep}. Similar to frequency selection based on domain knowledge, ECoG grid electrodes and fMRI voxels are often subselected by hand or with statistical tests. In general, these extracted features can then be put into almost any type of decoder, such as linear (or logistic) regression or a deep neural network (e.g.~\cite{ahmadi2019decoding}). 

It is also possible to let a deep learning architecture do more of the feature extraction. One approach is to first convert each electrode's signal into a frequency domain representation over time (i.e., a spectrogram), often via a wavelet transform. Then, this 2-dimensional representation (like an image) is provided as input to a CNN~\cite{golshan2020lfp,frey2019deepinsight,wang2019deep, barger2019robust} (Fig.~\ref{fig:2}B). If multiple electrode channels are being used for decoding, each channel can be fed into an independent CNN, or alternatively, the CNN weights for each channel can be shared \cite{frey2019deepinsight}. The CNN will then learn the relevant frequency domain representation for the decoding.

Another approach is to provide the raw input signals into a deep learning architecture (Fig.~\ref{fig:2}C). To learn temporal features, typically the signal is fed into a 1-dimensional CNN, where the convolutions occur in the time domain. This has been done with a standard CNN \cite{supratak2017deepsleepnet}, in addition to variant architectures. \citet{ahmadi2019end} used a temporal convolutional network, which is a more complex version of a 1-dimensional CNN that (among other things) allows for multiple timescales of inputs to affect the output. \citet{li2017targeting} used parameterized versions of temporal filters that target synchrony between electrodes. These convolutional approaches will automatically learn temporal filters (like frequency bands) that are relevant for decoding. 

In addition to temporal structure, there is often spatial structure of the electrode channels that can also be leveraged for decoding (Fig.~\ref{fig:2}A). Convolutional filters can be used in the spatial domain to learn spatial representations that are relevant for decoding, for example local functional correlation structure. It is common for the temporal filters and spatial filters to be learned in successive layers of the network, either temporal followed by spatial \cite{schirrmeister2017deep,lawhern2018eegnet} or vice-versa \cite{xie2018decoding}. Additionally, 3-dimensional convolutional filters can be learned that simultaneously incorporate both temporal and (2-dimensional) spatial dimensions \cite{angrick2019speech} or 3 spatial dimensions~\cite{zou20173d}. Including spatial filters, which is most common in EEG and ECoG, can help learn spatial motifs that are most relevant for the task. Moreover, from a practical perspective, convolutional networks are an efficient way of processing high-dimensional spatial data.

\section{The outputs of decoding}
Neural decoding is used to predict many outputs, including movement, speech, vision, and more. Sometimes, the output variable will be directly predicted from the neural inputs, e.g., when predicting movement velocities. Other times, the decoder may be trained to predict some intermediate representation, which has a predetermined mapping to the output (Fig.~\ref{fig:3}). For example, a GAN can be trained to generate an image using a small number of latent variables. This mapping from the low-dimensional variables to images can be learned without having to simultaneously record neural activity. Then, to decode an image from neural activity, one can train the decoder to predict the latent variables to be fed into the GAN, rather than the entire high-dimensional image. This two-step approach can be especially beneficial when the output data is complex and high-dimensional, as is often the case in vision or speech. In effect, the generative model can act as a prior on the underconstrained decoding solution. Across the following decoding outputs, researchers have used both the ``direct" and ``intermediate mapping" approaches (Fig.~\ref{fig:3}).

\subsection{Movement}
Some of the earliest uses of neural decoding were in the motor system \cite{georgopoulos1983spatial}. Researchers have used neural activity from motor cortex to predict many different motor outputs, such as movement kinematics (e.g., position and velocity), muscle activity (EMG), and broad type of movement. Traditionally, this decoding has used methods (e.g., Kalman Filter or Wiener Filter) that assumed a linear mapping from neural activity to the motor output, which has led to many successes \cite{wu2003neural,gilja2012high,serruya2002instant,carmena2003learning}. To improve the decoders, these methods were extended to allow specific nonlinearities (e.g., Unscented Kalman Filter and Wiener Cascade \cite{li2009unscented,luu2016gait,pohlmeyer2007prediction,ethier2012restoration}). Within the last decade, deep learning methods have become more common, frequently outperforming linear methods and their direct nonlinear extensions when compared (e.g., \cite{glaser2017machine,tseng2019decoding,sussillo2016making,xie2018decoding}).

\begin{figure}[!tpb]%figure3
\centering
\includegraphics[width=\textwidth]{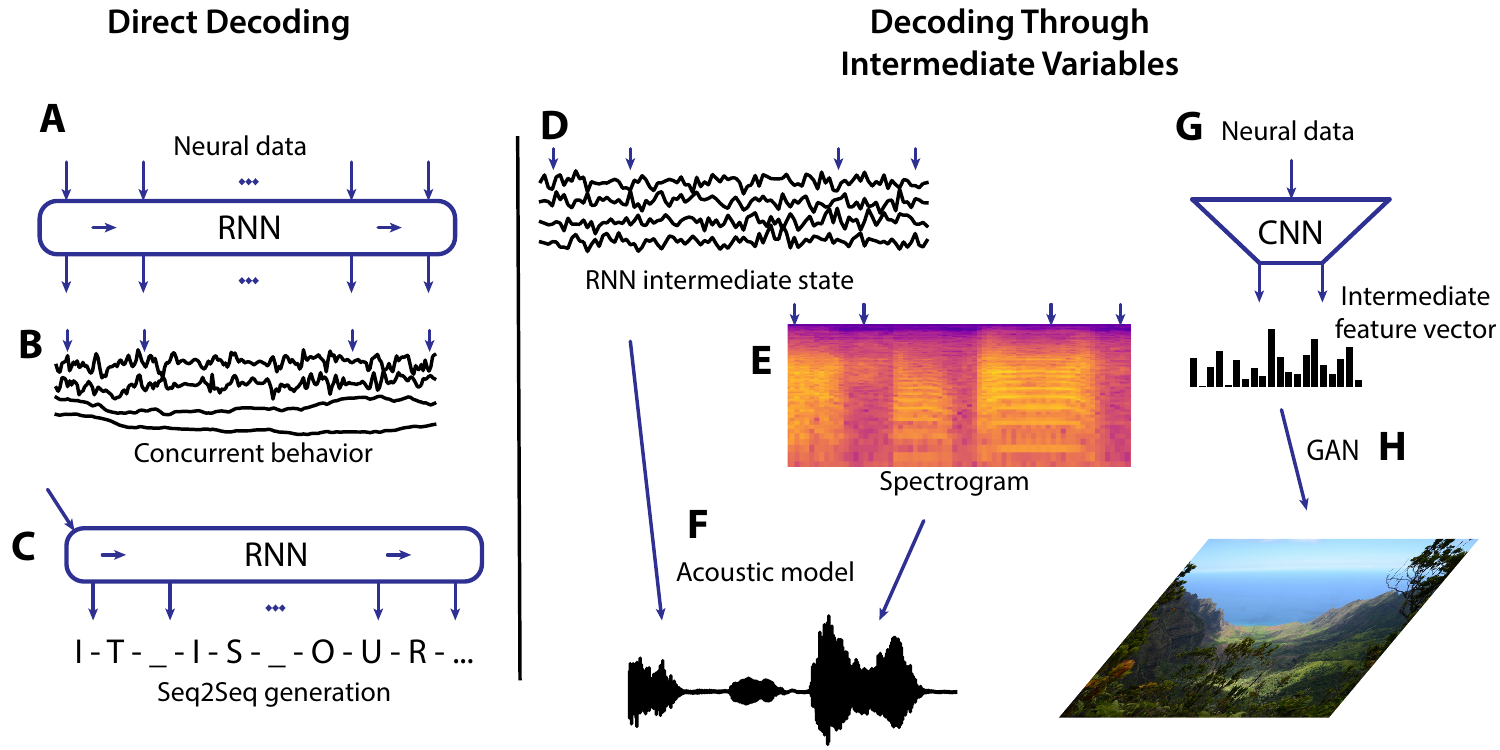}
\caption{Architectures and outputs of decoding. {\bf A:} Sequential inputs can be processed by RNNs which can use past context (or past and future in bi-directional RNNs). {\bf B:} RNN outputs at each timestep can be mapped to behaviors, e.g., movements, measured concurrently. {\bf C:} The final output of an RNN can be used as the input to a decoding network which can produce a second sequence of a different length, such as text. {\bf D:} RNNs can produce an intermediate state to be used in a second decoding step. {\bf E:} Intermediate states can often be structured, such as a spectrogram in this example. {\bf F:} Intermediate states can be fed into an acoustic model which produces acoustic waveforms. {\bf G:} Image-like inputs can be processed by CNNs to produce intermediate feature vectors. {\bf H:} Feature vectors can be fed into generative image models, e.g., a GAN, to produce a more realistic looking image.}\label{fig:3}
\end{figure}

Deep learning methods for decoding movement have been applied to a wide range of problems. Researchers have used many input signals that have high temporal resolution, including spikes \cite{naufel2019muscle,glaser2017machine,park2019estimation,wang2018decoding,sussillo2016making,tseng2019decoding,sussillo2012recurrent}, wide-band \cite{schwemmer2018meeting,skomrock2018characterization}, LFP \cite{ahmadi2019end,ahmadi2019decoding}, EEG \cite{nakagome2020empirical,nurse2016decoding}, and ECoG \cite{xie2018decoding,du2018decoding,pan2018rapid, elango2017sequence}. Additionally, deep learning has been used to predict many different outputs. Often the output is a continuous variable, such as the position, angle, or velocity of a limb, joint, or cursor \cite{glaser2017machine,wang2018decoding,sussillo2016making,nakagome2020empirical,tseng2019decoding,sussillo2012recurrent,ahmadi2019end,ahmadi2019decoding,xie2018decoding}, or a muscle’s EMG \cite{naufel2019muscle} (Fig.~\ref{fig:3}B). Rather than predicting a continuous variable, sometimes the goal is to classify different movement types \cite{du2018decoding,pan2018rapid,elango2017sequence,nurse2016decoding,schwemmer2018meeting,skomrock2018characterization}, for example, classifying which finger is moving \cite{du2018decoding}. Finally, deep learning decoders have been used to predict movements from effectors across different parts of the body, including arm  \cite{glaser2017machine,park2019estimation,sussillo2016making,tseng2019decoding,sussillo2012recurrent,ahmadi2019end,ahmadi2019decoding}, leg  \cite{wang2018decoding,nakagome2020empirical,tseng2019decoding}, wrist \cite{naufel2019muscle,schwemmer2018meeting,skomrock2018characterization}, and finger movements \cite{xie2018decoding,du2018decoding,pan2018rapid,elango2017sequence,schwemmer2018meeting,skomrock2018characterization}. Thus, deep learning methods have shown to be a very flexible tool for movement decoding.

RNNs are by far the most common deep learning architecture for movement decoding. When predicting a continuous movement variable, there is generally a linear mapping from the RNN’s output to the movement variable. When classifying movements, there is an additional softmax nonlinearity that determines the movement with the highest probability. From a deep learning perspective, given that this is a problem of converting one sequence (a temporal trace of neural activities) into another sequence (motor outputs), it would be expected that an RNN would be an appropriate architecture. Recurrent architectures also make sense from a scientific perspective:  motor cortical activity has dynamics that are important for producing movements \cite{shenoy2013cortical}, plus movements themselves have dynamics.

LSTMs have generally been the most common and successful type of RNN for decoding \cite{glaser2017machine,naufel2019muscle,park2019estimation,wang2018decoding,tseng2019decoding, ahmadi2019decoding, xie2018decoding, du2018decoding,pan2018rapid,elango2017sequence}, although other standard types of RNN architectures (e.g., GRUs \cite{nakagome2020empirical} and echostate networks \cite{sussillo2012recurrent}) have also proven successful. Additionally, researchers have found that stacking multiple layers of LSTMs \cite{tseng2019decoding, du2018decoding} can improve performance beyond a single LSTM \cite{tseng2019decoding}. LSTMs are likely successful because they are able to learn long-term dependencies better than a standard ``vanilla" RNN \cite{goodfellow2016deep}.

A common goal of neural decoding of movement is to be able to create a usable brain computer interface for patients. While the majority of deep learning uses have been in offline scenarios (decoding after the neural recording), there are several successful examples of real-time uses of deep learning for movement decoding \cite{sussillo2012recurrent, schwemmer2018meeting, skomrock2018characterization, sussillo2016making}. For example, in human patients with tetraplegia who had implanted electrode arrays, \citet{schwemmer2018meeting} were able to classify planned movements of wrist extension, wrist flexion, index extension, and index flexion. They then applied functional electrical stimulation to activate muscles according to this decoder, so that the patient was able to make these movements in real time. In \citet{sussillo2012recurrent}, monkeys with implanted electrode arrays were able to control the velocity of a cursor on a screen in real time.

While there has been great initial success, there are several challenges associated with using deep learning for real-time decoding for brain computer interfaces. One challenge is that the source of the recorded neural activity can change across days, for example due to slight movement of implanted electrodes. One approach that has dealt with this is the multiplicative RNN, which allows mappings from the neural input to the motor output to partially change across days \cite{sussillo2016making}. Another challenge is computation time, as there is the need to make predictions through the deep learning architecture at very high temporal resolution. When using a less complicated echostate network, \citet{sussillo2012recurrent} were able to decode with less than 25 ms temporal resolution. However, when using a more complex architecture of LSTMs followed by CNNs, \citet{schwemmer2018meeting} decoded at 100 ms resolution, slower than our perception. Relatedly, for linear methods that can be fit rapidly, researchers are able to adapt the decoder in real time to better match the subject’s intention (trying to get to a target) to improve performance \cite{gilja2012high,luu2016gait}. Developing similar approaches for deep learning based decoders is an exciting, unexplored area.

\subsection{Speech}
Vocal articulation is a complex behavior that engages a large functional area of the brain to produce movements that have a high degree of articulatory temporal and spatial precision~\cite{bouchard2013functional}. It is also a uniqely human ability which limits the recording modalities and neuroscientific interventions that can be used to study it. Due to the functional and temporal requirements of decoding speech, cortical surface electrical potentials recorded using ECoG is the typical recording modality used, although penetrating electrodes, MEG, EEG, and fNIRS are also used~\cite{chan2011decoding, herff2016automatic, sereshkeh2017eeg, wang2017towards}. When decoding from ECoG or EEG, researchers commonly use the signals' high gamma amplitude~\cite{bouchard2014neural}, although some use more broad spectrotemporal features as well~\cite{bouchard2014neural, mugler2014direct, akbari2019towards}.

Many approaches to decoding speech from neural signals have used some combination of linear methods and shallow probabilistic models. Clustering, SVMs, LDA, linear regression, and probabilistic models have been used with spectrotemporal features of electrical potentials to decode vowel acoustics, speech articulator movements, phonemes, whole words, and semantic categories~\cite{conant2018human, kellis2010decoding, chan2011decoding, mugler2014direct, bouchard2014neural, herff2015brain, guenther2009wireless}.

Deep learning approaches to decoding speech from neural signals have emerged that can potentially learn nonlinear mappings. Some of these approaches have operated on temporally segmented neural data and have thus used fully connected neural network architectures. For example, spectrotemporal features derived from ECoG or EEG have been used to reconstruct perceived spectrograms, classify words or syllables, or classify entire phrases~\cite{yang2015speech, sereshkeh2017eeg, livezey2019deep, wang2017towards, akbari2019towards}. These examples with temporally segmented neural data are useful for increasing understanding about neural representations, and as a step towards decoding natural speech. 

Mapping directly from continuous, time-varying neural signals to speech is the goal of speech brain-computer interfaces~\citep{wolpaw2002brain, schultz2017biosignal}. Both convolutional and recurrent networks are able to flexibly decode timeseries data and are often used for decoding naturalistic speech. \citet{heelan2019decoding} reconstructed perceived speech audio from multi-unit spike counts from a non-human primate and found that LSTM-based  networks outperformed other traditional and deep models. Speech represented as text does not have a simple one-to-one temporal alignment to regularly sampled neural signals. For this reason, speech-to-text decoding networks often use architectures and methods like sequence-to-sequence models or the connectionist temporal classification loss~\cite{sutskever2014sequence, graves2006connectionist}, which are commonly used in machine translation or automated speech recognition applications. As such, several groups have decoded directly from neural signals to text using recurrent networks such as sequence-to-sequence models~\cite{sun2019brain2char, makin2020machine} (Fig.~\ref{fig:3}C).

For decoding intelligible acoustic speech, it is also common to split decoding into a more constrained neural-to-intermediate mapping, followed by a second stage that maps this intermediate format into an acoustic waveform using acoustic priors for speech based on deep learning or hand-engineered methods. For instance, high gamma features recorded using ECoG have been used to decode spectrograms and speech articulator dynamics~\cite{angrick2019speech, anumanchipalli2019speech} as intermediate states. Then, either a WaveNet deep network~\cite{oord2016wavenet} was used to directly produce an acoustic waveform from the spectrogram~\cite{angrick2019speech}, or an RNN was used to produce acoustic features which were fed into a speech synthesizer~\cite{anumanchipalli2019speech}. These second stages do not require invasive neural data for training and were trained on a larger second corpus.

Deep learning models have improved the accuracy of primarily offline speech decoding tasks. Many of the preprocessing and decoding methods reviewed here are done offline using acausal or high-latency deep learning models. Developing deep learning methods, software, and hardware for real-time speech decoding is important for clinical applications of brain computer interfaces~\cite{guenther2009wireless, moses2019real}.

\subsection{Vision}

Similar to decoding acoustic speech, decoding visual stimuli from neural signals requires strong image priors due to the large variability of natural scenes and the relatively small bit-rate of neural recordings. Early attempts to reconstruct the full visual experience restricted decoding to simple images~\cite{miyawaki2008visual} or relied on a filterbank encoding model and a large set of natural images as a sampled prior~\cite{nishimoto2011reconstructing}. \citet{qiao2019category} solved the simpler task of classifying perceived object category using one CNN to select a small set of fMRI voxels which were fed into a second RNN for classification. Similarly, \citet{ellis2018high} classified among many visual scenes from calcium imaging data using feedforward or convolutional neural networks.  

As mentioned in \nameref{sec:architectures}, deep generative image models, such as GANs, can produce realistic images. In addition, CNNs trained to classify large naturalistic image databases~\cite{deng2009imagenet} (discriminative models) have been shown to encode a large amount of textural and semantic meaning in their activations~\cite{gatys2016image}, which can be used as an image prior. Due to the variety of ways that natural image priors can be created with deep networks, there exist decoding methods that combine different aspects of both generative and discriminative networks.

Given a deep generative model of images, a simpler decoder can be trained to map from neural data to the latent space of the model~\cite{seeliger2018generative, guccluturk2017reconstructing}, and the generative model can be used for image reconstruction. Similarly, a linear stage reconstruction followed by a deep network that cleans-up the image has been used with retinal ganglion cell output~\cite{parthasarathy2017neural}. Generative models can also be trained to reconstruct images directly from fMRI responses on real data with data augmentation from a simulated encoding model~\cite{st2018generative}.

Alternatively, generative and discriminative models can be used together. By leveraging a pretrained CNN, a simple decoder can be trained to map neural data to CNN activations that can then be passed into a convolutional image reconstruction model~\cite{wen2018neural}.  Additionally, the input image in a pretrained CNN can be optimized so that the CNN activations match predictions given by the fMRI responses~\cite{shen2019deep}. Researchers have also used an end-to-end approach in which they train the generative part directly on neural data with both an adversarial loss and a pretrained CNN feature loss~\cite{shen2019end}. Along with acoustic speech, decoding naturalistic visual stimuli presents one of the best cases to study the use of data-driven priors derived from deep networks.

\subsection{Other outputs}
While we have chosen to focus on a few decoding outputs that are prevalent in the literature, deep learning has been used for a myriad of decoding applications. RNNs such as LSTMs have been used to decode an animal’s location \cite{glaser2017machine,tampuu2019efficient,rezaei2018comparison,frey2019deepinsight} and direction \cite{xu2019comparison} from spiking activity in the hippocampus and head-direction cells, respectively. LSTMs have been used to decode what is being remembered in a working memory task from human fMRI \cite{li2019interpretable}. Researchers have used LSTMs \cite{yoo2018classification} and feedforward neural networks \cite{batty2019behavenet} to classify different classes of behaviors, using spiking activity in animals \cite{batty2019behavenet} and fNIRS measurements in humans \cite{yoo2018classification}. LSTMs \cite{hofmann2018decoding,garg2019merged} and CNNs \cite{tripathi2017using} have been used to classify emotions from EEG signals. Feedforward neural networks have been used to determine the source of a subject’s attention, using EEG in humans \cite{ciccarelli2019comparison, de2017machine} and spiking activity in monkeys \cite{astrand2014comparison}. CNNs \cite{supratak2017deepsleepnet, wang2019deep, barger2019robust}, along with LSTMs \cite{supratak2017deepsleepnet} have been used to predict a subject's stage of sleep from their EEG. For almost any behavioral signal that can be decoded, someone has tried to use deep learning.

\section{Discussion}

Deep learning is an attractive method for use in neural decoding because of its ability to learn complex, nonlinear transformations from data. In many of the examples above, deep networks can outperform linear or shallow methods even on relatively small datasets; however, examples exist where this is not the case, especially when using fMRI~\cite{schulz2019deep, 10.3389/fpsyt.2020.00440} or fNIRS data~\cite{hennrich2015investigating}. Relatedly, there are many times in which using hand-engineered features can outperform an end-to-end neural network that will learn the features. This is more likely with limited amounts of data, and also when there is strong prior knowledge about the relevant features. One general machine learning approach to efficiently use limited data is transfer learning, in which a neural network trained in one scenario (typically with more data) is used a separate scenario. This has been used in neural decoding to more effectively train decoders for new subjects \cite{elango2017sequence, makin2020machine} and for new predicted outputs \cite{schwemmer2018meeting}. As the capability to generate ever larger datasets develops with automated, long-term experimental setups for single animals~\cite{dhawale2017automated} and large scale recordings across multiple animals~\cite{allen}, deep learning is well poised to take advantage of this flood of data. As dataset sizes increase, this will also allow more features to be learned through data-driven network training rather than being selected by-hand.

Although deep learning will inevitably improve decoding accuracy as neuroscientists collect larger datasets, extracting scientific knowledge from trained networks is still an area of active research. That is, can we understand the transformations deep networks are learning? In computer vision, layers that include spatial attention~\cite{xu2015show} and methods for performing feature attribution~\cite{sundararajan2017axiomatic} have been developed to understand what parts of the input are important for prediction, although the latter are an active area of research~\cite{adebayo2018sanity}. These methods could be used to attribute what channels, neurons, or time-points are most salient for decoding~\cite{sundararajan2017axiomatic}. Additionally, there are methods for understanding deep network representations in computer vision that examine the representations networks have learned across layers~\cite{olah2018building, microscope}. Using these methods may help to understand the transformations that occur within neural decoders, however results may be sensitive to the decoder's architecture and not purely the data's structure. While deep learning interpretability methods are not commonly used on decoders trained on neural data, there are a few examples of networks that were built with interpretability in mind or were investigated after training~\cite{li2017targeting, schirrmeister2017deep, li2019interpretable, livezey2019deep}.

When interpreting decoders, it is often assumed that the decoder reveals the information contained in the brain about the decoded variable. It is important to note that this is only partially true when priors are being used for decoding \cite{kriegeskorte2019interpreting}, which is often the case when decoding a full image or acoustic speech. In these scenarios, the decoded outputs will be a function of both neural activity and the prior, so one cannot simply determine what information the brain has about the output.

The software used to create, train, and evaluate deep networks has been steadily developed and is now almost as easy to use as other standard machine learning methods. A wide range of cost functions, layer types, and parameter optimization algorithms are implemented and accessible in deep learning libraries such as PyTorch or Tensorflow~\cite{paszke2019pytorch, abadi2016tensorflow} and libraries in other programming languages. Like other machine learning methods, care must be taken to carefully cross-validate results as deep networks can easily overfit to the training data.

In addition to their use in neural decoding, deep learning has other prominent uses within neuroscience \cite{kietzmann2018deep, richards2019deep}. Neural networks have a long history in neuroscience as models of neural processing~\cite{hopfield1982neural, zipser1988back}. More recently, there has also been a surge of papers using deep networks as encoding models~\cite{sussillo2015neural, mcintosh2016deep, kell2018task}. There has been a specific focus on using the representations learned by deep networks trained to perform behavioral tasks (e.g., image recognition) to predict neural responses in corresponding brain areas (e.g., across the visual hierarchy~\cite{yamins2016using}). Combining these multiple complementary approaches is one promising approach to understanding neural computation.

\section*{Acknowledgements}
We would like to thank Ella Batty and Charles Frye for very helpful comments on this manuscript.

\section*{Funding}
JIG was supported by National Science Foundation NeuroNex Award DBI-1707398 and The Gatsby Foundation AT3708. JAL was supported by the LBNL Laboratory Directed Research and Development program.

\bibliography{refs_arxiv}

\begin{thebibliography}{140}
\providecommand{\natexlab}[1]{#1}
\providecommand{\url}[1]{\texttt{#1}}
\expandafter\ifx\csname urlstyle\endcsname\relax
  \providecommand{\doi}[1]{doi: #1}\else
  \providecommand{\doi}{doi: \begingroup \urlstyle{rm}\Url}\fi

\bibitem[Quiroga et~al.(2006)Quiroga, Snyder, Batista, Cui, and
  Andersen]{quiroga2006movement}
Rodrigo~Quian Quiroga, Lawrence~H Snyder, Aaron~P Batista, He~Cui, and
  Richard~A Andersen.
\newblock Movement intention is better predicted than attention in the
  posterior parietal cortex.
\newblock \emph{Journal of neuroscience}, 26\penalty0 (13):\penalty0
  3615--3620, 2006.

\bibitem[Harrison and Tong(2009)]{harrison2009decoding}
Stephenie~A Harrison and Frank Tong.
\newblock Decoding reveals the contents of visual working memory in early
  visual areas.
\newblock \emph{Nature}, 458\penalty0 (7238):\penalty0 632--635, 2009.

\bibitem[Acharya et~al.(2010)Acharya, Fifer, Benz, Crone, and
  Thakor]{acharya2010electrocorticographic}
Soumyadipta Acharya, Matthew~S Fifer, Heather~L Benz, Nathan~E Crone, and
  Nitish~V Thakor.
\newblock Electrocorticographic amplitude predicts finger positions during slow
  grasping motions of the hand.
\newblock \emph{Journal of neural engineering}, 7\penalty0 (4):\penalty0
  046002, 2010.

\bibitem[Weygandt et~al.(2012)Weygandt, Blecker, Sch{\"a}fer, Hackmack, Haynes,
  Vaitl, Stark, and Schienle]{weygandt2012fmri}
Martin Weygandt, Carlo~R Blecker, Axel Sch{\"a}fer, Kerstin Hackmack,
  John-Dylan Haynes, Dieter Vaitl, Rudolf Stark, and Anne Schienle.
\newblock fmri pattern recognition in obsessive--compulsive disorder.
\newblock \emph{Neuroimage}, 60\penalty0 (2):\penalty0 1186--1193, 2012.

\bibitem[Rich and Wallis(2016)]{rich2016decoding}
Erin~L Rich and Jonathan~D Wallis.
\newblock Decoding subjective decisions from orbitofrontal cortex.
\newblock \emph{Nature neuroscience}, 19\penalty0 (7):\penalty0 973, 2016.

\bibitem[Glaser et~al.(2018)Glaser, Perich, Ramkumar, Miller, and
  Kording]{glaser2018population}
Joshua~I Glaser, Matthew~G Perich, Pavan Ramkumar, Lee~E Miller, and Konrad~P
  Kording.
\newblock Population coding of conditional probability distributions in dorsal
  premotor cortex.
\newblock \emph{Nature communications}, 9\penalty0 (1):\penalty0 1--14, 2018.

\bibitem[Hamilton et~al.(2018)Hamilton, Edwards, and
  Chang]{hamilton2018spatial}
Liberty~S Hamilton, Erik Edwards, and Edward~F Chang.
\newblock A spatial map of onset and sustained responses to speech in the human
  superior temporal gyrus.
\newblock \emph{Current Biology}, 28\penalty0 (12):\penalty0 1860--1871, 2018.

\bibitem[Brackbill et~al.(2020)Brackbill, Rhoades, Kling, Shah, Sher, Litke,
  and Chichilnisky]{brackbill2020reconstruction}
Nora Brackbill, Colleen Rhoades, Alexandra Kling, Nishal~P Shah, Alexander
  Sher, Alan~M Litke, and EJ~Chichilnisky.
\newblock Reconstruction of natural images from responses of primate retinal
  ganglion cells.
\newblock \emph{bioRxiv}, 2020.

\bibitem[McIntosh et~al.(2016)McIntosh, Maheswaranathan, Nayebi, Ganguli, and
  Baccus]{mcintosh2016deep}
Lane McIntosh, Niru Maheswaranathan, Aran Nayebi, Surya Ganguli, and Stephen
  Baccus.
\newblock Deep learning models of the retinal response to natural scenes.
\newblock In \emph{Advances in neural information processing systems}, pages
  1369--1377, 2016.

\bibitem[Nagamine and Mesgarani(2017)]{nagamine2017understanding}
Tasha Nagamine and Nima Mesgarani.
\newblock Understanding the representation and computation of multilayer
  perceptrons: A case study in speech recognition.
\newblock In \emph{Proceedings of the 34th International Conference on Machine
  Learning-Volume 70}, pages 2564--2573. JMLR. org, 2017.

\bibitem[Kell et~al.(2018)Kell, Yamins, Shook, Norman-Haignere, and
  McDermott]{kell2018task}
Alexander~JE Kell, Daniel~LK Yamins, Erica~N Shook, Sam~V Norman-Haignere, and
  Josh~H McDermott.
\newblock A task-optimized neural network replicates human auditory behavior,
  predicts brain responses, and reveals a cortical processing hierarchy.
\newblock \emph{Neuron}, 98\penalty0 (3):\penalty0 630--644, 2018.

\bibitem[Livezey et~al.(2019)Livezey, Bouchard, and Chang]{livezey2019deep}
Jesse~A Livezey, Kristofer~E Bouchard, and Edward~F Chang.
\newblock Deep learning as a tool for neural data analysis: speech
  classification and cross-frequency coupling in human sensorimotor cortex.
\newblock \emph{PLoS computational biology}, 15\penalty0 (9):\penalty0
  e1007091, 2019.

\bibitem[Alipanahi et~al.(2015)Alipanahi, Delong, Weirauch, and
  Frey]{alipanahi2015predicting}
Babak Alipanahi, Andrew Delong, Matthew~T Weirauch, and Brendan~J Frey.
\newblock Predicting the sequence specificities of dna-and rna-binding proteins
  by deep learning.
\newblock \emph{Nature biotechnology}, 33\penalty0 (8):\penalty0 831--838,
  2015.

\bibitem[Piech et~al.(2015)Piech, Bassen, Huang, Ganguli, Sahami, Guibas, and
  Sohl-Dickstein]{piech2015deep}
Chris Piech, Jonathan Bassen, Jonathan Huang, Surya Ganguli, Mehran Sahami,
  Leonidas~J Guibas, and Jascha Sohl-Dickstein.
\newblock Deep knowledge tracing.
\newblock In \emph{Advances in neural information processing systems}, pages
  505--513, 2015.

\bibitem[Paganini et~al.(2018)Paganini, de~Oliveira, and
  Nachman]{paganini2018calogan}
Michela Paganini, Luke de~Oliveira, and Benjamin Nachman.
\newblock Calogan: Simulating 3d high energy particle showers in multilayer
  electromagnetic calorimeters with generative adversarial networks.
\newblock \emph{Physical Review D}, 97\penalty0 (1):\penalty0 014021, 2018.

\bibitem[Kurth et~al.(2018)Kurth, Treichler, Romero, Mudigonda, Luehr,
  Phillips, Mahesh, Matheson, Deslippe, Fatica, et~al.]{kurth2018exascale}
Thorsten Kurth, Sean Treichler, Joshua Romero, Mayur Mudigonda, Nathan Luehr,
  Everett Phillips, Ankur Mahesh, Michael Matheson, Jack Deslippe, Massimiliano
  Fatica, et~al.
\newblock Exascale deep learning for climate analytics.
\newblock In \emph{SC18: International Conference for High Performance
  Computing, Networking, Storage and Analysis}, pages 649--660. IEEE, 2018.

\bibitem[Sch{\"u}tt et~al.(2018)Sch{\"u}tt, Sauceda, Kindermans, Tkatchenko,
  and M{\"u}ller]{schutt2018schnet}
Kristof~T Sch{\"u}tt, Huziel~E Sauceda, P-J Kindermans, Alexandre Tkatchenko,
  and K-R M{\"u}ller.
\newblock Schnet--a deep learning architecture for molecules and materials.
\newblock \emph{The Journal of Chemical Physics}, 148\penalty0 (24):\penalty0
  241722, 2018.

\bibitem[Hochreiter and Schmidhuber(1997)]{hochreiter1997long}
Sepp Hochreiter and J{\"u}rgen Schmidhuber.
\newblock Long short-term memory.
\newblock \emph{Neural computation}, 9\penalty0 (8):\penalty0 1735--1780, 1997.

\bibitem[Krizhevsky et~al.(2012)Krizhevsky, Sutskever, and
  Hinton]{krizhevsky2012imagenet}
Alex Krizhevsky, Ilya Sutskever, and Geoffrey~E Hinton.
\newblock Imagenet classification with deep convolutional neural networks.
\newblock In \emph{Advances in neural information processing systems}, pages
  1097--1105, 2012.

\bibitem[Sutskever et~al.(2014)Sutskever, Vinyals, and
  Le]{sutskever2014sequence}
Ilya Sutskever, Oriol Vinyals, and Quoc~V Le.
\newblock Sequence to sequence learning with neural networks.
\newblock In \emph{Advances in neural information processing systems}, pages
  3104--3112, 2014.

\bibitem[He et~al.(2016)He, Zhang, Ren, and Sun]{he2016deep}
Kaiming He, Xiangyu Zhang, Shaoqing Ren, and Jian Sun.
\newblock Deep residual learning for image recognition.
\newblock In \emph{Proceedings of the IEEE conference on computer vision and
  pattern recognition}, pages 770--778, 2016.

\bibitem[Amodei et~al.(2016)Amodei, Ananthanarayanan, Anubhai, Bai, Battenberg,
  Case, Casper, Catanzaro, Cheng, Chen, et~al.]{amodei2016deep}
Dario Amodei, Sundaram Ananthanarayanan, Rishita Anubhai, Jingliang Bai, Eric
  Battenberg, Carl Case, Jared Casper, Bryan Catanzaro, Qiang Cheng, Guoliang
  Chen, et~al.
\newblock Deep speech 2: End-to-end speech recognition in english and mandarin.
\newblock In \emph{International conference on machine learning}, pages
  173--182, 2016.

\bibitem[Wu et~al.(2020)Wu, Pan, Chen, Long, Zhang, and
  Philip]{wu2020comprehensive}
Zonghan Wu, Shirui Pan, Fengwen Chen, Guodong Long, Chengqi Zhang, and S~Yu
  Philip.
\newblock A comprehensive survey on graph neural networks.
\newblock \emph{IEEE Transactions on Neural Networks and Learning Systems},
  2020.

\bibitem[Vaswani et~al.(2017)Vaswani, Shazeer, Parmar, Uszkoreit, Jones, Gomez,
  Kaiser, and Polosukhin]{vaswani2017attention}
Ashish Vaswani, Noam Shazeer, Niki Parmar, Jakob Uszkoreit, Llion Jones,
  Aidan~N Gomez, {\L}ukasz Kaiser, and Illia Polosukhin.
\newblock Attention is all you need.
\newblock In \emph{Advances in neural information processing systems}, pages
  5998--6008, 2017.

\bibitem[Goodfellow et~al.(2014)Goodfellow, Pouget-Abadie, Mirza, Xu,
  Warde-Farley, Ozair, Courville, and Bengio]{goodfellow2014generative}
Ian Goodfellow, Jean Pouget-Abadie, Mehdi Mirza, Bing Xu, David Warde-Farley,
  Sherjil Ozair, Aaron Courville, and Yoshua Bengio.
\newblock Generative adversarial nets.
\newblock In \emph{Advances in neural information processing systems}, pages
  2672--2680, 2014.

\bibitem[Radford et~al.(2015)Radford, Metz, and
  Chintala]{radford2015unsupervised}
Alec Radford, Luke Metz, and Soumith Chintala.
\newblock Unsupervised representation learning with deep convolutional
  generative adversarial networks.
\newblock \emph{arXiv preprint arXiv:1511.06434}, 2015.

\bibitem[Parthasarathy et~al.(2017)Parthasarathy, Batty, Falcon, Rutten,
  Rajpal, Chichilnisky, and Paninski]{parthasarathy2017neural}
Nikhil Parthasarathy, Eleanor Batty, William Falcon, Thomas Rutten, Mohit
  Rajpal, EJ~Chichilnisky, and Liam Paninski.
\newblock Neural networks for efficient bayesian decoding of natural images
  from retinal neurons.
\newblock In \emph{Advances in Neural Information Processing Systems}, pages
  6434--6445, 2017.

\bibitem[Glaser et~al.(2017)Glaser, Chowdhury, Perich, Miller, and
  Kording]{glaser2017machine}
Joshua~I Glaser, Raeed~H Chowdhury, Matthew~G Perich, Lee~E Miller, and
  Konrad~P Kording.
\newblock Machine learning for neural decoding.
\newblock \emph{arXiv preprint arXiv:1708.00909}, 2017.

\bibitem[Bouchard and Chang(2019)]{bouchard_chang_2019}
Kristofer~E. Bouchard and Edward~F Chang.
\newblock Human ecog speaking consonant-vowel syllables, 2019.
\newblock URL \url{https://doi.org/10.6084/m9.figshare.c.4617263.v4}.

\bibitem[Kazemifar et~al.(2017)Kazemifar, Manning, Rajakumar, Gomez, Soddu,
  Borrie, Menon, Bartha, Initiative, et~al.]{kazemifar2017spontaneous}
Samaneh Kazemifar, Kathryn~Y Manning, Nagalingam Rajakumar, Francisco~A Gomez,
  Andrea Soddu, Michael~J Borrie, Ravi~S Menon, Robert Bartha, Alzheimer’s
  Disease~Neuroimaging Initiative, et~al.
\newblock Spontaneous low frequency bold signal variations from resting-state
  fmri are decreased in alzheimer disease.
\newblock \emph{PloS one}, 12\penalty0 (6), 2017.

\bibitem[Goodfellow et~al.(2016)Goodfellow, Bengio, and
  Courville]{goodfellow2016deep}
Ian Goodfellow, Yoshua Bengio, and Aaron Courville.
\newblock \emph{Deep learning}.
\newblock MIT press, 2016.

\bibitem[Buzs{\'a}ki et~al.(2012)Buzs{\'a}ki, Anastassiou, and
  Koch]{buzsaki2012origin}
Gy{\"o}rgy Buzs{\'a}ki, Costas~A Anastassiou, and Christof Koch.
\newblock The origin of extracellular fields and currents—eeg, ecog, lfp and
  spikes.
\newblock \emph{Nature reviews neuroscience}, 13\penalty0 (6):\penalty0
  407--420, 2012.

\bibitem[Chen et~al.(2013)Chen, Wardill, Sun, Pulver, Renninger, Baohan,
  Schreiter, Kerr, Orger, Jayaraman, et~al.]{chen2013ultrasensitive}
Tsai-Wen Chen, Trevor~J Wardill, Yi~Sun, Stefan~R Pulver, Sabine~L Renninger,
  Amy Baohan, Eric~R Schreiter, Rex~A Kerr, Michael~B Orger, Vivek Jayaraman,
  et~al.
\newblock Ultrasensitive fluorescent proteins for imaging neuronal activity.
\newblock \emph{Nature}, 499\penalty0 (7458):\penalty0 295--300, 2013.

\bibitem[Flint et~al.(2012)Flint, Ethier, Oby, Miller, and
  Slutzky]{flint2012local}
Robert~D Flint, Christian Ethier, Emily~R Oby, Lee~E Miller, and Marc~W
  Slutzky.
\newblock Local field potentials allow accurate decoding of muscle activity.
\newblock \emph{Journal of neurophysiology}, 108\penalty0 (1):\penalty0 18--24,
  2012.

\bibitem[Frey et~al.(2019)Frey, Tanni, Perrodin, O’Leary, Nau, Kelly, Banino,
  Doeller, and Barry]{frey2019deepinsight}
Markus Frey, Sander Tanni, Catherine Perrodin, Alice O’Leary, Matthias Nau,
  Jack Kelly, Andrea Banino, Christian~F Doeller, and Caswell Barry.
\newblock Deepinsight: a general framework for interpreting wide-band neural
  activity.
\newblock \emph{bioRxiv}, page 871848, 2019.

\bibitem[Maia~Chagas et~al.(2013)Maia~Chagas, Theis, Sengupta, St{\"u}ttgen,
  Bethge, and Schwarz]{maia2013functional}
Andr{\'e} Maia~Chagas, Lucas Theis, Biswa Sengupta, Maik~Christopher
  St{\"u}ttgen, Matthias Bethge, and Cornelius Schwarz.
\newblock Functional analysis of ultra high information rates conveyed by rat
  vibrissal primary afferents.
\newblock \emph{Frontiers in neural circuits}, 7:\penalty0 190, 2013.

\bibitem[Giovannucci et~al.(2019)Giovannucci, Friedrich, Gunn, Kalfon, Brown,
  Koay, Taxidis, Najafi, Gauthier, Zhou, et~al.]{giovannucci2019caiman}
Andrea Giovannucci, Johannes Friedrich, Pat Gunn, Jeremie Kalfon, Brandon~L
  Brown, Sue~Ann Koay, Jiannis Taxidis, Farzaneh Najafi, Jeffrey~L Gauthier,
  Pengcheng Zhou, et~al.
\newblock Caiman an open source tool for scalable calcium imaging data
  analysis.
\newblock \emph{Elife}, 8:\penalty0 e38173, 2019.

\bibitem[Vogelstein et~al.(2010)Vogelstein, Packer, Machado, Sippy, Babadi,
  Yuste, and Paninski]{vogelstein2010fast}
Joshua~T Vogelstein, Adam~M Packer, Timothy~A Machado, Tanya Sippy, Baktash
  Babadi, Rafael Yuste, and Liam Paninski.
\newblock Fast nonnegative deconvolution for spike train inference from
  population calcium imaging.
\newblock \emph{Journal of neurophysiology}, 104\penalty0 (6):\penalty0
  3691--3704, 2010.

\bibitem[Soltanian-Zadeh et~al.(2019)Soltanian-Zadeh, Sahingur, Blau, Gong, and
  Farsiu]{soltanian2019fast}
Somayyeh Soltanian-Zadeh, Kaan Sahingur, Sarah Blau, Yiyang Gong, and Sina
  Farsiu.
\newblock Fast and robust active neuron segmentation in two-photon calcium
  imaging using spatiotemporal deep learning.
\newblock \emph{Proceedings of the National Academy of Sciences}, 116\penalty0
  (17):\penalty0 8554--8563, 2019.

\bibitem[Speiser et~al.(2017)Speiser, Yan, Archer, Buesing, Turaga, and
  Macke]{speiser2017fast}
Artur Speiser, Jinyao Yan, Evan~W Archer, Lars Buesing, Srinivas~C Turaga, and
  Jakob~H Macke.
\newblock Fast amortized inference of neural activity from calcium imaging data
  with variational autoencoders.
\newblock In \emph{Advances in Neural Information Processing Systems}, pages
  4024--4034, 2017.

\bibitem[Bouchard and Chang(2014)]{bouchard2014neural}
Kristofer~E Bouchard and Edward~F Chang.
\newblock Neural decoding of spoken vowels from human sensory-motor cortex with
  high-density electrocorticography.
\newblock In \emph{2014 36th Annual International Conference of the IEEE
  Engineering in Medicine and Biology Society}, pages 6782--6785. IEEE, 2014.

\bibitem[Yang et~al.(2015)Yang, Sheth, Schevon, Ii, and
  Mesgarani]{yang2015speech}
Minda Yang, Sameer~A Sheth, Catherine~A Schevon, Guy M~Mckhann Ii, and Nima
  Mesgarani.
\newblock Speech reconstruction from human auditory cortex with deep neural
  networks.
\newblock In \emph{Sixteenth Annual Conference of the International Speech
  Communication Association}, 2015.

\bibitem[Mugler et~al.(2014)Mugler, Patton, Flint, Wright, Schuele, Rosenow,
  Shih, Krusienski, and Slutzky]{mugler2014direct}
Emily~M Mugler, James~L Patton, Robert~D Flint, Zachary~A Wright, Stephan~U
  Schuele, Joshua Rosenow, Jerry~J Shih, Dean~J Krusienski, and Marc~W Slutzky.
\newblock Direct classification of all american english phonemes using signals
  from functional speech motor cortex.
\newblock \emph{Journal of neural engineering}, 11\penalty0 (3):\penalty0
  035015, 2014.

\bibitem[Ahmadi et~al.(2019{\natexlab{a}})Ahmadi, Constandinou, and
  Bouganis]{ahmadi2019decoding}
Nur Ahmadi, Timothy~G Constandinou, and Christos-Savvas Bouganis.
\newblock Decoding hand kinematics from local field potentials using long
  short-term memory (lstm) network.
\newblock In \emph{2019 9th International IEEE/EMBS Conference on Neural
  Engineering (NER)}, pages 415--419. IEEE, 2019{\natexlab{a}}.

\bibitem[Golshan et~al.(2020)Golshan, Hebb, and Mahoor]{golshan2020lfp}
Hosein~M Golshan, Adam~O Hebb, and Mohammad~H Mahoor.
\newblock Lfp-net: A deep learning framework to recognize human behavioral
  activities using brain stn-lfp signals.
\newblock \emph{Journal of Neuroscience Methods}, 335:\penalty0 108621, 2020.

\bibitem[Wang et~al.(2019)Wang, Zhang, Ma, Huang, and Hong]{wang2019deep}
Jialin Wang, Yanchun Zhang, Qinying Ma, Huihui Huang, and Xiaoyuan Hong.
\newblock Deep learning for single-channel eeg signals sleep stage scoring
  based on frequency domain representation.
\newblock In \emph{International Conference on Health Information Science},
  pages 121--133. Springer, 2019.

\bibitem[Barger et~al.(2019)Barger, Frye, Liu, Dan, and
  Bouchard]{barger2019robust}
Zeke Barger, Charles~G Frye, Danqian Liu, Yang Dan, and Kristofer~E Bouchard.
\newblock Robust, automated sleep scoring by a compact neural network with
  distributional shift correction.
\newblock \emph{PloS one}, 14\penalty0 (12), 2019.

\bibitem[Supratak et~al.(2017)Supratak, Dong, Wu, and
  Guo]{supratak2017deepsleepnet}
Akara Supratak, Hao Dong, Chao Wu, and Yike Guo.
\newblock Deepsleepnet: a model for automatic sleep stage scoring based on raw
  single-channel eeg.
\newblock \emph{IEEE Transactions on Neural Systems and Rehabilitation
  Engineering}, 25\penalty0 (11):\penalty0 1998--2008, 2017.

\bibitem[Ahmadi et~al.(2019{\natexlab{b}})Ahmadi, Constandinou, and
  Bouganis]{ahmadi2019end}
Nur Ahmadi, Timothy~G Constandinou, and Christos-Savvas Bouganis.
\newblock End-to-end hand kinematic decoding from lfps using temporal
  convolutional network.
\newblock In \emph{2019 IEEE Biomedical Circuits and Systems Conference
  (BioCAS)}, pages 1--4. IEEE, 2019{\natexlab{b}}.

\bibitem[Li et~al.(2017)Li, Dzirasa, Carin, Carlson, et~al.]{li2017targeting}
Yitong Li, Kafui Dzirasa, Lawrence Carin, David~E Carlson, et~al.
\newblock Targeting eeg/lfp synchrony with neural nets.
\newblock In \emph{Advances in Neural Information Processing Systems}, pages
  4620--4630, 2017.

\bibitem[Schirrmeister et~al.(2017)Schirrmeister, Springenberg, Fiederer,
  Glasstetter, Eggensperger, Tangermann, Hutter, Burgard, and
  Ball]{schirrmeister2017deep}
Robin~Tibor Schirrmeister, Jost~Tobias Springenberg, Lukas Dominique~Josef
  Fiederer, Martin Glasstetter, Katharina Eggensperger, Michael Tangermann,
  Frank Hutter, Wolfram Burgard, and Tonio Ball.
\newblock Deep learning with convolutional neural networks for eeg decoding and
  visualization.
\newblock \emph{Human brain mapping}, 38\penalty0 (11):\penalty0 5391--5420,
  2017.

\bibitem[Lawhern et~al.(2018)Lawhern, Solon, Waytowich, Gordon, Hung, and
  Lance]{lawhern2018eegnet}
Vernon~J Lawhern, Amelia~J Solon, Nicholas~R Waytowich, Stephen~M Gordon,
  Chou~P Hung, and Brent~J Lance.
\newblock Eegnet: a compact convolutional neural network for eeg-based
  brain--computer interfaces.
\newblock \emph{Journal of neural engineering}, 15\penalty0 (5):\penalty0
  056013, 2018.

\bibitem[Xie et~al.(2018)Xie, Schwartz, and Prasad]{xie2018decoding}
Ziqian Xie, Odelia Schwartz, and Abhishek Prasad.
\newblock Decoding of finger trajectory from ecog using deep learning.
\newblock \emph{Journal of neural engineering}, 15\penalty0 (3):\penalty0
  036009, 2018.

\bibitem[Angrick et~al.(2019)Angrick, Herff, Mugler, Tate, Slutzky, Krusienski,
  and Schultz]{angrick2019speech}
Miguel Angrick, Christian Herff, Emily Mugler, Matthew~C Tate, Marc~W Slutzky,
  Dean~J Krusienski, and Tanja Schultz.
\newblock Speech synthesis from ecog using densely connected 3d convolutional
  neural networks.
\newblock \emph{Journal of neural engineering}, 16\penalty0 (3):\penalty0
  036019, 2019.

\bibitem[Zou et~al.(2017)Zou, Zheng, Miao, Mckeown, and Wang]{zou20173d}
Liang Zou, Jiannan Zheng, Chunyan Miao, Martin~J Mckeown, and Z~Jane Wang.
\newblock 3d cnn based automatic diagnosis of attention deficit hyperactivity
  disorder using functional and structural mri.
\newblock \emph{IEEE Access}, 5:\penalty0 23626--23636, 2017.

\bibitem[Georgopoulos et~al.(1983)Georgopoulos, Caminiti, Kalaska, and
  Massey]{georgopoulos1983spatial}
Apostolos~P Georgopoulos, Roberto Caminiti, John~F Kalaska, and Joseph~T
  Massey.
\newblock Spatial coding of movement: a hypothesis concerning the coding of
  movement direction by motor cortical populations.
\newblock \emph{Experimental Brain Research}, 49\penalty0 (Suppl. 7):\penalty0
  327--336, 1983.

\bibitem[Wu et~al.(2003)Wu, Black, Gao, Serruya, Shaikhouni, Donoghue, and
  Bienenstock]{wu2003neural}
Wei Wu, Michael~J Black, Yun Gao, M~Serruya, A~Shaikhouni, JP~Donoghue, and
  Elie Bienenstock.
\newblock Neural decoding of cursor motion using a kalman filter.
\newblock In \emph{Advances in neural information processing systems}, pages
  133--140, 2003.

\bibitem[Gilja et~al.(2012)Gilja, Nuyujukian, Chestek, Cunningham, Byron, Fan,
  Churchland, Kaufman, Kao, Ryu, et~al.]{gilja2012high}
Vikash Gilja, Paul Nuyujukian, Cindy~A Chestek, John~P Cunningham, M~Yu Byron,
  Joline~M Fan, Mark~M Churchland, Matthew~T Kaufman, Jonathan~C Kao, Stephen~I
  Ryu, et~al.
\newblock A high-performance neural prosthesis enabled by control algorithm
  design.
\newblock \emph{Nature neuroscience}, 15\penalty0 (12):\penalty0 1752, 2012.

\bibitem[Serruya et~al.(2002)Serruya, Hatsopoulos, Paninski, Fellows, and
  Donoghue]{serruya2002instant}
Mijail~D Serruya, Nicholas~G Hatsopoulos, Liam Paninski, Matthew~R Fellows, and
  John~P Donoghue.
\newblock Instant neural control of a movement signal.
\newblock \emph{Nature}, 416\penalty0 (6877):\penalty0 141--142, 2002.

\bibitem[Carmena et~al.(2003)Carmena, Lebedev, Crist, O'Doherty, Santucci,
  Dimitrov, Patil, Henriquez, and Nicolelis]{carmena2003learning}
Jose~M Carmena, Mikhail~A Lebedev, Roy~E Crist, Joseph~E O'Doherty, David~M
  Santucci, Dragan~F Dimitrov, Parag~G Patil, Craig~S Henriquez, and Miguel~AL
  Nicolelis.
\newblock Learning to control a brain--machine interface for reaching and
  grasping by primates.
\newblock \emph{PLoS biology}, 1\penalty0 (2), 2003.

\bibitem[Li et~al.(2009)Li, O'Doherty, Hanson, Lebedev, Henriquez, and
  Nicolelis]{li2009unscented}
Zheng Li, Joseph~E O'Doherty, Timothy~L Hanson, Mikhail~A Lebedev, Craig~S
  Henriquez, and Miguel~AL Nicolelis.
\newblock Unscented kalman filter for brain-machine interfaces.
\newblock \emph{PloS one}, 4\penalty0 (7), 2009.

\bibitem[Luu et~al.(2016)Luu, He, Brown, Nakagome, and
  Contreras-Vidal]{luu2016gait}
Trieu~Phat Luu, Yongtian He, Samuel Brown, Sho Nakagome, and Jose~L
  Contreras-Vidal.
\newblock Gait adaptation to visual kinematic perturbations using a real-time
  closed-loop brain--computer interface to a virtual reality avatar.
\newblock \emph{Journal of neural engineering}, 13\penalty0 (3):\penalty0
  036006, 2016.

\bibitem[Pohlmeyer et~al.(2007)Pohlmeyer, Solla, Perreault, and
  Miller]{pohlmeyer2007prediction}
Eric~A Pohlmeyer, Sara~A Solla, Eric~J Perreault, and Lee~E Miller.
\newblock Prediction of upper limb muscle activity from motor cortical
  discharge during reaching.
\newblock \emph{Journal of neural engineering}, 4\penalty0 (4):\penalty0 369,
  2007.

\bibitem[Ethier et~al.(2012)Ethier, Oby, Bauman, and
  Miller]{ethier2012restoration}
Christian Ethier, Emily~R Oby, Matthew~J Bauman, and Lee~E Miller.
\newblock Restoration of grasp following paralysis through brain-controlled
  stimulation of muscles.
\newblock \emph{Nature}, 485\penalty0 (7398):\penalty0 368--371, 2012.

\bibitem[Tseng et~al.(2019)Tseng, Urpi, Lebedev, and
  Nicolelis]{tseng2019decoding}
Po-He Tseng, N{\'u}ria~Armengol Urpi, Mikhail Lebedev, and Miguel Nicolelis.
\newblock Decoding movements from cortical ensemble activity using a long
  short-term memory recurrent network.
\newblock \emph{Neural computation}, 31\penalty0 (6):\penalty0 1085--1113,
  2019.

\bibitem[Sussillo et~al.(2016)Sussillo, Stavisky, Kao, Ryu, and
  Shenoy]{sussillo2016making}
David Sussillo, Sergey~D Stavisky, Jonathan~C Kao, Stephen~I Ryu, and Krishna~V
  Shenoy.
\newblock Making brain--machine interfaces robust to future neural variability.
\newblock \emph{Nature communications}, 7:\penalty0 13749, 2016.

\bibitem[Naufel et~al.(2019)Naufel, Glaser, Kording, Perreault, and
  Miller]{naufel2019muscle}
Stephanie Naufel, Joshua~I Glaser, Konrad~P Kording, Eric~J Perreault, and
  Lee~E Miller.
\newblock A muscle-activity-dependent gain between motor cortex and emg.
\newblock \emph{Journal of neurophysiology}, 121\penalty0 (1):\penalty0 61--73,
  2019.

\bibitem[Park and Kim(2019)]{park2019estimation}
Jisung Park and Sung-Phil Kim.
\newblock Estimation of speed and direction of arm movements from m1 activity
  using a nonlinear neural decoder.
\newblock In \emph{2019 7th International Winter Conference on Brain-Computer
  Interface (BCI)}, pages 1--4. IEEE, 2019.

\bibitem[Wang et~al.(2018)Wang, Truccolo, and Borton]{wang2018decoding}
Yinong Wang, Wilson Truccolo, and David~A Borton.
\newblock Decoding hindlimb kinematics from primate motor cortex using long
  short-term memory recurrent neural networks.
\newblock In \emph{2018 40th Annual International Conference of the IEEE
  Engineering in Medicine and Biology Society (EMBC)}, pages 1944--1947. IEEE,
  2018.

\bibitem[Sussillo et~al.(2012)Sussillo, Nuyujukian, Fan, Kao, Stavisky, Ryu,
  and Shenoy]{sussillo2012recurrent}
David Sussillo, Paul Nuyujukian, Joline~M Fan, Jonathan~C Kao, Sergey~D
  Stavisky, Stephen Ryu, and Krishna Shenoy.
\newblock A recurrent neural network for closed-loop intracortical
  brain--machine interface decoders.
\newblock \emph{Journal of neural engineering}, 9\penalty0 (2):\penalty0
  026027, 2012.

\bibitem[Schwemmer et~al.(2018)Schwemmer, Skomrock, Sederberg, Ting, Sharma,
  Bockbrader, and Friedenberg]{schwemmer2018meeting}
Michael~A Schwemmer, Nicholas~D Skomrock, Per~B Sederberg, Jordyn~E Ting,
  Gaurav Sharma, Marcia~A Bockbrader, and David~A Friedenberg.
\newblock Meeting brain--computer interface user performance expectations using
  a deep neural network decoding framework.
\newblock \emph{Nature medicine}, 24\penalty0 (11):\penalty0 1669--1676, 2018.

\bibitem[Skomrock et~al.(2018)Skomrock, Schwemmer, Ting, Trivedi, Sharma,
  Bockbrader, and Friedenberg]{skomrock2018characterization}
Nicholas~D Skomrock, Michael~A Schwemmer, Jordyn~E Ting, Hemang~R Trivedi,
  Gaurav Sharma, Marcia~A Bockbrader, and David~A Friedenberg.
\newblock A characterization of brain-computer interface performance trade-offs
  using support vector machines and deep neural networks to decode movement
  intent.
\newblock \emph{Frontiers in neuroscience}, 12:\penalty0 763, 2018.

\bibitem[Nakagome et~al.(2020)Nakagome, Luu, He, Ravindran, and
  Contreras-Vidal]{nakagome2020empirical}
Sho Nakagome, Trieu~Phat Luu, Yongtian He, Akshay~Sujatha Ravindran, and Jose~L
  Contreras-Vidal.
\newblock An empirical comparison of neural networks and machine learning
  algorithms for eeg gait decoding.
\newblock \emph{Scientific Reports}, 10\penalty0 (1):\penalty0 1--17, 2020.

\bibitem[Nurse et~al.(2016)Nurse, Mashford, Yepes, Kiral-Kornek, Harrer, and
  Freestone]{nurse2016decoding}
Ewan Nurse, Benjamin~S Mashford, Antonio~Jimeno Yepes, Isabell Kiral-Kornek,
  Stefan Harrer, and Dean~R Freestone.
\newblock Decoding eeg and lfp signals using deep learning: heading truenorth.
\newblock In \emph{Proceedings of the ACM International Conference on Computing
  Frontiers}, pages 259--266, 2016.

\bibitem[Du et~al.(2018)Du, Yang, Liu, and Huang]{du2018decoding}
Anming Du, Shuqin Yang, Weijia Liu, and Haiping Huang.
\newblock Decoding ecog signal with deep learning model based on lstm.
\newblock In \emph{TENCON 2018-2018 IEEE Region 10 Conference}, pages
  0430--0435. IEEE, 2018.

\bibitem[Pan et~al.(2018)Pan, Li, Qi, Yu, Zhu, Zheng, Wang, and
  Zhang]{pan2018rapid}
Gang Pan, Jia-Jun Li, Yu~Qi, Hang Yu, Jun-Ming Zhu, Xiao-Xiang Zheng, Yue-Ming
  Wang, and Shao-Min Zhang.
\newblock Rapid decoding of hand gestures in electrocorticography using
  recurrent neural networks.
\newblock \emph{Frontiers in neuroscience}, 12:\penalty0 555, 2018.

\bibitem[Elango et~al.(2017)Elango, Patel, Miller, and
  Gilja]{elango2017sequence}
Venkatesh Elango, Aashish~N Patel, Kai~J Miller, and Vikash Gilja.
\newblock Sequence transfer learning for neural decoding.
\newblock \emph{bioRxiv}, page 210732, 2017.

\bibitem[Shenoy et~al.(2013)Shenoy, Sahani, and Churchland]{shenoy2013cortical}
Krishna~V Shenoy, Maneesh Sahani, and Mark~M Churchland.
\newblock Cortical control of arm movements: a dynamical systems perspective.
\newblock \emph{Annual review of neuroscience}, 36:\penalty0 337--359, 2013.

\bibitem[Bouchard et~al.(2013)Bouchard, Mesgarani, Johnson, and
  Chang]{bouchard2013functional}
Kristofer~E Bouchard, Nima Mesgarani, Keith Johnson, and Edward~F Chang.
\newblock Functional organization of human sensorimotor cortex for speech
  articulation.
\newblock \emph{Nature}, 495\penalty0 (7441):\penalty0 327, 2013.

\bibitem[Chan et~al.(2011)Chan, Halgren, Marinkovic, and
  Cash]{chan2011decoding}
Alexander~M Chan, Eric Halgren, Ksenija Marinkovic, and Sydney~S Cash.
\newblock Decoding word and category-specific spatiotemporal representations
  from meg and eeg.
\newblock \emph{Neuroimage}, 54\penalty0 (4):\penalty0 3028--3039, 2011.

\bibitem[Herff and Schultz(2016)]{herff2016automatic}
Christian Herff and Tanja Schultz.
\newblock Automatic speech recognition from neural signals: a focused review.
\newblock \emph{Frontiers in neuroscience}, 10:\penalty0 429, 2016.

\bibitem[Sereshkeh et~al.(2017)Sereshkeh, Trott, Bricout, and
  Chau]{sereshkeh2017eeg}
Alborz~Rezazadeh Sereshkeh, Robert Trott, Aur{\'e}lien Bricout, and Tom Chau.
\newblock Eeg classification of covert speech using regularized neural
  networks.
\newblock \emph{IEEE/ACM Transactions on Audio, Speech, and Language
  Processing}, 25\penalty0 (12):\penalty0 2292--2300, 2017.

\bibitem[Wang et~al.(2017)Wang, Kim, Hernandez-Mulero, Heitzman, and
  Ferrari]{wang2017towards}
Jun Wang, Myungjong Kim, Angel~W Hernandez-Mulero, Daragh Heitzman, and Paul
  Ferrari.
\newblock Towards decoding speech production from single-trial
  magnetoencephalography (meg) signals.
\newblock In \emph{2017 IEEE International Conference on Acoustics, Speech and
  Signal Processing (ICASSP)}, pages 3036--3040. IEEE, 2017.

\bibitem[Akbari et~al.(2019)Akbari, Khalighinejad, Herrero, Mehta, and
  Mesgarani]{akbari2019towards}
Hassan Akbari, Bahar Khalighinejad, Jose~L Herrero, Ashesh~D Mehta, and Nima
  Mesgarani.
\newblock Towards reconstructing intelligible speech from the human auditory
  cortex.
\newblock \emph{Scientific reports}, 9\penalty0 (1):\penalty0 1--12, 2019.

\bibitem[Conant et~al.(2018)Conant, Bouchard, Leonard, and
  Chang]{conant2018human}
David~F Conant, Kristofer~E Bouchard, Matthew~K Leonard, and Edward~F Chang.
\newblock Human sensorimotor cortex control of directly measured vocal tract
  movements during vowel production.
\newblock \emph{Journal of Neuroscience}, 38\penalty0 (12):\penalty0
  2955--2966, 2018.

\bibitem[Kellis et~al.(2010)Kellis, Miller, Thomson, Brown, House, and
  Greger]{kellis2010decoding}
Spencer Kellis, Kai Miller, Kyle Thomson, Richard Brown, Paul House, and
  Bradley Greger.
\newblock Decoding spoken words using local field potentials recorded from the
  cortical surface.
\newblock \emph{Journal of neural engineering}, 7\penalty0 (5):\penalty0
  056007, 2010.

\bibitem[Herff et~al.(2015)Herff, Heger, De~Pesters, Telaar, Brunner, Schalk,
  and Schultz]{herff2015brain}
Christian Herff, Dominic Heger, Adriana De~Pesters, Dominic Telaar, Peter
  Brunner, Gerwin Schalk, and Tanja Schultz.
\newblock Brain-to-text: decoding spoken phrases from phone representations in
  the brain.
\newblock \emph{Frontiers in neuroscience}, 9:\penalty0 217, 2015.

\bibitem[Guenther et~al.(2009)Guenther, Brumberg, Wright, Nieto-Castanon,
  Tourville, Panko, Law, Siebert, Bartels, Andreasen,
  et~al.]{guenther2009wireless}
Frank~H Guenther, Jonathan~S Brumberg, E~Joseph Wright, Alfonso Nieto-Castanon,
  Jason~A Tourville, Mikhail Panko, Robert Law, Steven~A Siebert, Jess~L
  Bartels, Dinal~S Andreasen, et~al.
\newblock A wireless brain-machine interface for real-time speech synthesis.
\newblock \emph{PloS one}, 4\penalty0 (12), 2009.

\bibitem[Wolpaw et~al.(2002)Wolpaw, Birbaumer, McFarland, Pfurtscheller, and
  Vaughan]{wolpaw2002brain}
Jonathan~R Wolpaw, Niels Birbaumer, Dennis~J McFarland, Gert Pfurtscheller, and
  Theresa~M Vaughan.
\newblock Brain--computer interfaces for communication and control.
\newblock \emph{Clinical neurophysiology}, 113\penalty0 (6):\penalty0 767--791,
  2002.

\bibitem[Schultz et~al.(2017)Schultz, Wand, Hueber, Krusienski, Herff, and
  Brumberg]{schultz2017biosignal}
Tanja Schultz, Michael Wand, Thomas Hueber, Dean~J Krusienski, Christian Herff,
  and Jonathan~S Brumberg.
\newblock Biosignal-based spoken communication: A survey.
\newblock \emph{IEEE/ACM Transactions on Audio, Speech, and Language
  Processing}, 25\penalty0 (12):\penalty0 2257--2271, 2017.

\bibitem[Heelan et~al.(2019)Heelan, Lee, O’Shea, Lynch, Brandman, Truccolo,
  and Nurmikko]{heelan2019decoding}
Christopher Heelan, Jihun Lee, Ronan O’Shea, Laurie Lynch, David~M Brandman,
  Wilson Truccolo, and Arto~V Nurmikko.
\newblock Decoding speech from spike-based neural population recordings in
  secondary auditory cortex of non-human primates.
\newblock \emph{Communications biology}, 2\penalty0 (1):\penalty0 1--12, 2019.

\bibitem[Graves et~al.(2006)Graves, Fern{\'a}ndez, Gomez, and
  Schmidhuber]{graves2006connectionist}
Alex Graves, Santiago Fern{\'a}ndez, Faustino Gomez, and J{\"u}rgen
  Schmidhuber.
\newblock Connectionist temporal classification: labelling unsegmented sequence
  data with recurrent neural networks.
\newblock In \emph{Proceedings of the 23rd international conference on Machine
  learning}, pages 369--376, 2006.

\bibitem[Sun et~al.(2019)Sun, Anumanchipalli, and Chang]{sun2019brain2char}
Pengfei Sun, Gopala~K Anumanchipalli, and Edward~F Chang.
\newblock Brain2char: A deep architecture for decoding text from brain
  recordings.
\newblock \emph{arXiv preprint arXiv:1909.01401}, 2019.

\bibitem[Makin et~al.(2020)Makin, Moses, and Chang]{makin2020machine}
Joseph~G Makin, David~A Moses, and Edward~F Chang.
\newblock Machine translation of cortical activity to text with an
  encoder--decoder framework.
\newblock Technical report, Nature Publishing Group, 2020.

\bibitem[Anumanchipalli et~al.(2019)Anumanchipalli, Chartier, and
  Chang]{anumanchipalli2019speech}
Gopala~K Anumanchipalli, Josh Chartier, and Edward~F Chang.
\newblock Speech synthesis from neural decoding of spoken sentences.
\newblock \emph{Nature}, 568\penalty0 (7753):\penalty0 493, 2019.

\bibitem[Oord et~al.(2016)Oord, Dieleman, Zen, Simonyan, Vinyals, Graves,
  Kalchbrenner, Senior, and Kavukcuoglu]{oord2016wavenet}
Aaron van~den Oord, Sander Dieleman, Heiga Zen, Karen Simonyan, Oriol Vinyals,
  Alex Graves, Nal Kalchbrenner, Andrew Senior, and Koray Kavukcuoglu.
\newblock Wavenet: A generative model for raw audio.
\newblock \emph{arXiv preprint arXiv:1609.03499}, 2016.

\bibitem[Moses et~al.(2019)Moses, Leonard, Makin, and Chang]{moses2019real}
David~A Moses, Matthew~K Leonard, Joseph~G Makin, and Edward~F Chang.
\newblock Real-time decoding of question-and-answer speech dialogue using human
  cortical activity.
\newblock \emph{Nature communications}, 10\penalty0 (1):\penalty0 1--14, 2019.

\bibitem[Miyawaki et~al.(2008)Miyawaki, Uchida, Yamashita, Sato, Morito,
  Tanabe, Sadato, and Kamitani]{miyawaki2008visual}
Yoichi Miyawaki, Hajime Uchida, Okito Yamashita, Masa-aki Sato, Yusuke Morito,
  Hiroki~C Tanabe, Norihiro Sadato, and Yukiyasu Kamitani.
\newblock Visual image reconstruction from human brain activity using a
  combination of multiscale local image decoders.
\newblock \emph{Neuron}, 60\penalty0 (5):\penalty0 915--929, 2008.

\bibitem[Nishimoto et~al.(2011)Nishimoto, Vu, Naselaris, Benjamini, Yu, and
  Gallant]{nishimoto2011reconstructing}
Shinji Nishimoto, An~T Vu, Thomas Naselaris, Yuval Benjamini, Bin Yu, and
  Jack~L Gallant.
\newblock Reconstructing visual experiences from brain activity evoked by
  natural movies.
\newblock \emph{Current Biology}, 21\penalty0 (19):\penalty0 1641--1646, 2011.

\bibitem[Qiao et~al.(2019)Qiao, Chen, Wang, Zhang, Zeng, Tong, and
  Yan]{qiao2019category}
Kai Qiao, Jian Chen, Linyuan Wang, Chi Zhang, Lei Zeng, Li~Tong, and Bin Yan.
\newblock Category decoding of visual stimuli from human brain activity using a
  bidirectional recurrent neural network to simulate bidirectional information
  flows in human visual cortices.
\newblock \emph{Frontiers in neuroscience}, 13, 2019.

\bibitem[Ellis and Michaelides(2018)]{ellis2018high}
Randall~Jordan Ellis and Michael Michaelides.
\newblock High-accuracy decoding of complex visual scenes from neuronal calcium
  responses.
\newblock \emph{BioRxiv}, page 271296, 2018.

\bibitem[Deng et~al.(2009)Deng, Dong, Socher, Li, Li, and
  Fei-Fei]{deng2009imagenet}
Jia Deng, Wei Dong, Richard Socher, Li-Jia Li, Kai Li, and Li~Fei-Fei.
\newblock Imagenet: A large-scale hierarchical image database.
\newblock In \emph{2009 IEEE conference on computer vision and pattern
  recognition}, pages 248--255. Ieee, 2009.

\bibitem[Gatys et~al.(2016)Gatys, Ecker, and Bethge]{gatys2016image}
Leon~A Gatys, Alexander~S Ecker, and Matthias Bethge.
\newblock Image style transfer using convolutional neural networks.
\newblock In \emph{Proceedings of the IEEE conference on computer vision and
  pattern recognition}, pages 2414--2423, 2016.

\bibitem[Seeliger et~al.(2018)Seeliger, G{\"u}{\c{c}}l{\"u}, Ambrogioni,
  G{\"u}{\c{c}}l{\"u}t{\"u}rk, and van Gerven]{seeliger2018generative}
Katja Seeliger, Umut G{\"u}{\c{c}}l{\"u}, Luca Ambrogioni, Yagmur
  G{\"u}{\c{c}}l{\"u}t{\"u}rk, and Marcel~AJ van Gerven.
\newblock Generative adversarial networks for reconstructing natural images
  from brain activity.
\newblock \emph{NeuroImage}, 181:\penalty0 775--785, 2018.

\bibitem[G{\"u}{\c{c}}l{\"u}t{\"u}rk et~al.(2017)G{\"u}{\c{c}}l{\"u}t{\"u}rk,
  G{\"u}{\c{c}}l{\"u}, Seeliger, Bosch, van Lier, and van
  Gerven]{guccluturk2017reconstructing}
Ya{\u{g}}mur G{\"u}{\c{c}}l{\"u}t{\"u}rk, Umut G{\"u}{\c{c}}l{\"u}, Katja
  Seeliger, Sander Bosch, Rob van Lier, and Marcel~AJ van Gerven.
\newblock Reconstructing perceived faces from brain activations with deep
  adversarial neural decoding.
\newblock In \emph{Advances in Neural Information Processing Systems}, pages
  4246--4257, 2017.

\bibitem[St-Yves and Naselaris(2018)]{st2018generative}
Ghislain St-Yves and Thomas Naselaris.
\newblock Generative adversarial networks conditioned on brain activity
  reconstruct seen images.
\newblock In \emph{2018 IEEE International Conference on Systems, Man, and
  Cybernetics (SMC)}, pages 1054--1061. IEEE, 2018.

\bibitem[Wen et~al.(2018)Wen, Shi, Zhang, Lu, Cao, and Liu]{wen2018neural}
Haiguang Wen, Junxing Shi, Yizhen Zhang, Kun-Han Lu, Jiayue Cao, and Zhongming
  Liu.
\newblock Neural encoding and decoding with deep learning for dynamic natural
  vision.
\newblock \emph{Cerebral Cortex}, 28\penalty0 (12):\penalty0 4136--4160, 2018.

\bibitem[Shen et~al.(2019{\natexlab{a}})Shen, Horikawa, Majima, and
  Kamitani]{shen2019deep}
Guohua Shen, Tomoyasu Horikawa, Kei Majima, and Yukiyasu Kamitani.
\newblock Deep image reconstruction from human brain activity.
\newblock \emph{PLoS computational biology}, 15\penalty0 (1):\penalty0
  e1006633, 2019{\natexlab{a}}.

\bibitem[Shen et~al.(2019{\natexlab{b}})Shen, Dwivedi, Majima, Horikawa, and
  Kamitani]{shen2019end}
Guohua Shen, Kshitij Dwivedi, Kei Majima, Tomoyasu Horikawa, and Yukiyasu
  Kamitani.
\newblock End-to-end deep image reconstruction from human brain activity.
\newblock \emph{Frontiers in Computational Neuroscience}, 13,
  2019{\natexlab{b}}.

\bibitem[Tampuu et~al.(2019)Tampuu, Matiisen, {\'O}lafsd{\'o}ttir, Barry, and
  Vicente]{tampuu2019efficient}
Ardi Tampuu, Tambet Matiisen, H~Freyja {\'O}lafsd{\'o}ttir, Caswell Barry, and
  Raul Vicente.
\newblock Efficient neural decoding of self-location with a deep recurrent
  network.
\newblock \emph{PLoS computational biology}, 15\penalty0 (2):\penalty0
  e1006822, 2019.

\bibitem[Rezaei et~al.(2018)Rezaei, Gillespie, Guidera, Nazari, Sadri, Frank,
  Eden, and Yousefi]{rezaei2018comparison}
Mohammad~R Rezaei, Anna~K Gillespie, Jennifer~A Guidera, Behzad Nazari, Saeid
  Sadri, Loren~M Frank, Uri~T Eden, and Ali Yousefi.
\newblock A comparison study of point-process filter and deep learning
  performance in estimating rat position using an ensemble of place cells.
\newblock In \emph{2018 40th Annual International Conference of the IEEE
  Engineering in Medicine and Biology Society (EMBC)}, pages 4732--4735. IEEE,
  2018.

\bibitem[Xu et~al.(2019)Xu, Wu, Winter, Mehlman, Butler, Simmons, Harvey,
  Berkowitz, Chen, Taube, et~al.]{xu2019comparison}
Zishen Xu, Wei Wu, Shawn~S Winter, Max~L Mehlman, William~N Butler, Christine~M
  Simmons, Ryan~E Harvey, Laura~E Berkowitz, Yang Chen, Jeffrey~S Taube, et~al.
\newblock A comparison of neural decoding methods and population coding across
  thalamo-cortical head direction cells.
\newblock \emph{Frontiers in Neural Circuits}, 13, 2019.

\bibitem[Li and Fan(2019)]{li2019interpretable}
Hongming Li and Yong Fan.
\newblock Interpretable, highly accurate brain decoding of subtly distinct
  brain states from functional mri using intrinsic functional networks and long
  short-term memory recurrent neural networks.
\newblock \emph{NeuroImage}, 202:\penalty0 116059, 2019.

\bibitem[Yoo et~al.(2018)Yoo, Woo, and Amad]{yoo2018classification}
So-Hyeon Yoo, Seong-Woo Woo, and Zafar Amad.
\newblock Classification of three categories from prefrontal cortex using lstm
  networks: fnirs study.
\newblock In \emph{2018 18th International Conference on Control, Automation
  and Systems (ICCAS)}, pages 1141--1146. IEEE, 2018.

\bibitem[Batty et~al.(2019)Batty, Whiteway, Saxena, Biderman, Abe, Musall,
  Gillis, Markowitz, Churchland, Cunningham, et~al.]{batty2019behavenet}
Eleanor Batty, Matthew Whiteway, Shreya Saxena, Dan Biderman, Taiga Abe, Simon
  Musall, Winthrop Gillis, Jeffrey Markowitz, Anne Churchland, John~P
  Cunningham, et~al.
\newblock Behavenet: nonlinear embedding and bayesian neural decoding of
  behavioral videos.
\newblock In \emph{Advances in Neural Information Processing Systems}, pages
  15680--15691, 2019.

\bibitem[Hofmann et~al.(2018)Hofmann, Klotzsche, Mariola, Nikulin, Villringer,
  and Gaebler]{hofmann2018decoding}
Simon~M Hofmann, Felix Klotzsche, Alberto Mariola, Vadim~V Nikulin, Arno
  Villringer, and Michael Gaebler.
\newblock Decoding subjective emotional arousal during a naturalistic vr
  experience from eeg using lstms.
\newblock In \emph{2018 IEEE International Conference on Artificial
  Intelligence and Virtual Reality (AIVR)}, pages 128--131. IEEE, 2018.

\bibitem[Garg et~al.(2019)Garg, Kapoor, Bedi, and Sunkaria]{garg2019merged}
Anumit Garg, Ashna Kapoor, Anterpreet~Kaur Bedi, and Ramesh~K Sunkaria.
\newblock Merged lstm model for emotion classification using eeg signals.
\newblock In \emph{2019 International Conference on Data Science and
  Engineering (ICDSE)}, pages 139--143. IEEE, 2019.

\bibitem[Tripathi et~al.(2017)Tripathi, Acharya, Sharma, Mittal, and
  Bhattacharya]{tripathi2017using}
Samarth Tripathi, Shrinivas Acharya, Ranti~Dev Sharma, Sudhanshu Mittal, and
  Samit Bhattacharya.
\newblock Using deep and convolutional neural networks for accurate emotion
  classification on deap dataset.
\newblock In \emph{Twenty-Ninth IAAI Conference}, 2017.

\bibitem[Ciccarelli et~al.(2019)Ciccarelli, Nolan, Perricone, Calamia, Haro,
  O’Sullivan, Mesgarani, Quatieri, and Smalt]{ciccarelli2019comparison}
Gregory Ciccarelli, Michael Nolan, Joseph Perricone, Paul~T Calamia, Stephanie
  Haro, James O’Sullivan, Nima Mesgarani, Thomas~F Quatieri, and
  Christopher~J Smalt.
\newblock Comparison of two-talker attention decoding from eeg with nonlinear
  neural networks and linear methods.
\newblock \emph{Scientific reports}, 9\penalty0 (1):\penalty0 1--10, 2019.

\bibitem[de~Taillez et~al.(2017)de~Taillez, Kollmeier, and
  Meyer]{de2017machine}
Tobias de~Taillez, Birger Kollmeier, and Bernd~T Meyer.
\newblock Machine learning for decoding listeners’ attention from
  electroencephalography evoked by continuous speech.
\newblock \emph{European Journal of Neuroscience}, 2017.

\bibitem[Astrand et~al.(2014)Astrand, Enel, Ibos, Dominey, Baraduc, and
  Hamed]{astrand2014comparison}
Elaine Astrand, Pierre Enel, Guilhem Ibos, Peter~Ford Dominey, Pierre Baraduc,
  and Suliann~Ben Hamed.
\newblock Comparison of classifiers for decoding sensory and cognitive
  information from prefrontal neuronal populations.
\newblock \emph{PloS one}, 9\penalty0 (1), 2014.

\bibitem[Schulz et~al.(2019)Schulz, Yeo, Vogelstein, Mourao-Miranada, Kather,
  Kording, Richards, and Bzdok]{schulz2019deep}
Marc-Andre Schulz, Thomas Yeo, Joshua Vogelstein, Janaina Mourao-Miranada,
  Jakob Kather, Konrad Kording, Blake~A Richards, and Danilo Bzdok.
\newblock Deep learning for brains?: Different linear and nonlinear scaling in
  uk biobank brain images vs. machine-learning datasets.
\newblock \emph{bioRxiv}, page 757054, 2019.

\bibitem[Thomas et~al.(2020)Thomas, Gallo, Cerliani, Zhutovsky, El-Gazzar, and
  van Wingen]{10.3389/fpsyt.2020.00440}
Rajat~Mani Thomas, Selene Gallo, Leonardo Cerliani, Paul Zhutovsky, Ahmed
  El-Gazzar, and Guido van Wingen.
\newblock Classifying autism spectrum disorder using the temporal statistics of
  resting-state functional mri data with 3d convolutional neural networks.
\newblock \emph{Frontiers in Psychiatry}, 11:\penalty0 440, 2020.

\bibitem[Hennrich et~al.(2015)Hennrich, Herff, Heger, and
  Schultz]{hennrich2015investigating}
Johannes Hennrich, Christian Herff, Dominic Heger, and Tanja Schultz.
\newblock Investigating deep learning for fnirs based bci.
\newblock In \emph{2015 37th Annual international conference of the IEEE
  Engineering in Medicine and Biology Society (EMBC)}, pages 2844--2847. IEEE,
  2015.

\bibitem[Dhawale et~al.(2017)Dhawale, Poddar, Wolff, Normand, Kopelowitz, and
  {\"O}lveczky]{dhawale2017automated}
Ashesh~K Dhawale, Rajesh Poddar, Steffen~BE Wolff, Valentin~A Normand, Evi
  Kopelowitz, and Bence~P {\"O}lveczky.
\newblock Automated long-term recording and analysis of neural activity in
  behaving animals.
\newblock \emph{Elife}, 6:\penalty0 e27702, 2017.

\bibitem[Observatory(2016)]{allen}
Allen~Brain Observatory.
\newblock Available at: http://observatory.brain-map.org/visualcoding, 2016.

\bibitem[Xu et~al.(2015)Xu, Ba, Kiros, Cho, Courville, Salakhudinov, Zemel, and
  Bengio]{xu2015show}
Kelvin Xu, Jimmy Ba, Ryan Kiros, Kyunghyun Cho, Aaron Courville, Ruslan
  Salakhudinov, Rich Zemel, and Yoshua Bengio.
\newblock Show, attend and tell: Neural image caption generation with visual
  attention.
\newblock In \emph{International conference on machine learning}, pages
  2048--2057, 2015.

\bibitem[Sundararajan et~al.(2017)Sundararajan, Taly, and
  Yan]{sundararajan2017axiomatic}
Mukund Sundararajan, Ankur Taly, and Qiqi Yan.
\newblock Axiomatic attribution for deep networks.
\newblock In \emph{Proceedings of the 34th International Conference on Machine
  Learning-Volume 70}, pages 3319--3328. JMLR. org, 2017.

\bibitem[Adebayo et~al.(2018)Adebayo, Gilmer, Muelly, Goodfellow, Hardt, and
  Kim]{adebayo2018sanity}
Julius Adebayo, Justin Gilmer, Michael Muelly, Ian Goodfellow, Moritz Hardt,
  and Been Kim.
\newblock Sanity checks for saliency maps.
\newblock In \emph{Advances in Neural Information Processing Systems}, pages
  9505--9515, 2018.

\bibitem[Olah et~al.(2018)Olah, Satyanarayan, Johnson, Carter, Schubert, Ye,
  and Mordvintsev]{olah2018building}
Chris Olah, Arvind Satyanarayan, Ian Johnson, Shan Carter, Ludwig Schubert,
  Katherine Ye, and Alexander Mordvintsev.
\newblock The building blocks of interpretability.
\newblock \emph{Distill}, 3\penalty0 (3):\penalty0 e10, 2018.

\bibitem[mic(2020)]{microscope}
The {OpenAI} microscope.
\newblock https://microscope.openai.com/models, 2020.
\newblock Accessed: 2020-05-12.

\bibitem[Kriegeskorte and Douglas(2019)]{kriegeskorte2019interpreting}
Nikolaus Kriegeskorte and Pamela~K Douglas.
\newblock Interpreting encoding and decoding models.
\newblock \emph{Current opinion in neurobiology}, 55:\penalty0 167--179, 2019.

\bibitem[Paszke et~al.(2019)Paszke, Gross, Massa, Lerer, Bradbury, Chanan,
  Killeen, Lin, Gimelshein, Antiga, et~al.]{paszke2019pytorch}
Adam Paszke, Sam Gross, Francisco Massa, Adam Lerer, James Bradbury, Gregory
  Chanan, Trevor Killeen, Zeming Lin, Natalia Gimelshein, Luca Antiga, et~al.
\newblock Pytorch: An imperative style, high-performance deep learning library.
\newblock In \emph{Advances in Neural Information Processing Systems}, pages
  8024--8035, 2019.

\bibitem[Abadi et~al.(2016)Abadi, Barham, Chen, Chen, Davis, Dean, Devin,
  Ghemawat, Irving, Isard, et~al.]{abadi2016tensorflow}
Mart{\'\i}n Abadi, Paul Barham, Jianmin Chen, Zhifeng Chen, Andy Davis, Jeffrey
  Dean, Matthieu Devin, Sanjay Ghemawat, Geoffrey Irving, Michael Isard, et~al.
\newblock Tensorflow: A system for large-scale machine learning.
\newblock In \emph{12th $\{$USENIX$\}$ Symposium on Operating Systems Design
  and Implementation ($\{$OSDI$\}$ 16)}, pages 265--283, 2016.

\bibitem[Kietzmann et~al.(2018)Kietzmann, McClure, and
  Kriegeskorte]{kietzmann2018deep}
Tim~Christian Kietzmann, Patrick McClure, and Nikolaus Kriegeskorte.
\newblock Deep neural networks in computational neuroscience.
\newblock \emph{BioRxiv}, page 133504, 2018.

\bibitem[Richards et~al.(2019)Richards, Lillicrap, Beaudoin, Bengio, Bogacz,
  Christensen, Clopath, Costa, de~Berker, Ganguli, et~al.]{richards2019deep}
Blake~A Richards, Timothy~P Lillicrap, Philippe Beaudoin, Yoshua Bengio, Rafal
  Bogacz, Amelia Christensen, Claudia Clopath, Rui~Ponte Costa, Archy
  de~Berker, Surya Ganguli, et~al.
\newblock A deep learning framework for neuroscience.
\newblock \emph{Nature neuroscience}, 22\penalty0 (11):\penalty0 1761--1770,
  2019.

\bibitem[Hopfield(1982)]{hopfield1982neural}
John~J Hopfield.
\newblock Neural networks and physical systems with emergent collective
  computational abilities.
\newblock \emph{Proceedings of the national academy of sciences}, 79\penalty0
  (8):\penalty0 2554--2558, 1982.

\bibitem[Zipser and Andersen(1988)]{zipser1988back}
David Zipser and Richard~A Andersen.
\newblock A back-propagation programmed network that simulates response
  properties of a subset of posterior parietal neurons.
\newblock \emph{Nature}, 331\penalty0 (6158):\penalty0 679--684, 1988.

\bibitem[Sussillo et~al.(2015)Sussillo, Churchland, Kaufman, and
  Shenoy]{sussillo2015neural}
David Sussillo, Mark~M Churchland, Matthew~T Kaufman, and Krishna~V Shenoy.
\newblock A neural network that finds a naturalistic solution for the
  production of muscle activity.
\newblock \emph{Nature neuroscience}, 18\penalty0 (7):\penalty0 1025--1033,
  2015.

\bibitem[Yamins and DiCarlo(2016)]{yamins2016using}
Daniel~LK Yamins and James~J DiCarlo.
\newblock Using goal-driven deep learning models to understand sensory cortex.
\newblock \emph{Nature neuroscience}, 19\penalty0 (3):\penalty0 356, 2016.

\end{thebibliography}

% \begin{thebibliography}{0}

% \bibitem[Bofelli {\it et~al}., 2000]{Boffelli03}
% Bofelli,F., Name2, Name3 (2003) Article title, {\it Journal Name}, {\bf 199}, 133-154.

% \bibitem[Bag {\it et~al}., 2001]{Bag01}
% Bag,M., Name2, Name3 (2001) Article title, {\it Journal Name}, {\bf 99}, 33-54.

% \bibitem[Yoo \textit{et~al}., 2003]{Yoo03}
% Yoo,M.S. \textit{et~al}. (2003) Oxidative stress regulated genes
% in nigral dopaminergic neurnol cell: correlation with the known
% pathology in Parkinson's disease. \textit{Brain Res. Mol. Brain
% Res.}, \textbf{110}(Suppl. 1), 76--84.

% \bibitem[Lehmann, 1986]{Leh86}
% Lehmann,E.L. (1986) Chapter title. \textit{Book Title}. Vol.~1, 2nd edn. Springer-Verlag, New York.

% \bibitem[Crenshaw and Jones, 2003]{Cre03}
% Crenshaw, B.,III, and Jones, W.B.,Jr (2003) The future of clinical
% cancer management: one tumor, one chip. \textit{Bioinformatics},
% doi:10.1093/bioinformatics/btn000.

% \bibitem[Auhtor \textit{et~al}. (2000)]{Aut00}
% Auhtor,A.B. \textit{et~al}. (2000) Chapter title. In Smith, A.C.
% (ed.), \textit{Book Title}, 2nd edn. Publisher, Location, Vol. 1, pp.
% ???--???.

% \bibitem[Bardet, 1920]{Bar20}
% Bardet, G. (1920) Sur un syndrome d'obesite infantile avec
% polydactylie et retinite pigmentaire (contribution a l'etude des
% formes cliniques de l'obesite hypophysaire). PhD Thesis, name of
% institution, Paris, France.

% \end{thebibliography}
\end{document}